\documentclass[12pt,a4paper,final]{iopart}
\usepackage{iopams} 
\usepackage{graphicx}
\usepackage[breaklinks=true,colorlinks=true,linkcolor=blue,urlcolor=blue,citecolor=blue]{hyperref}

\usepackage{cite}

\begin{document}

\title[Defect physics in...]{Defect physics in complex energy materials}

\author{Khang Hoang$^{1,\ast}$ and M D Johannes$^2$}
\address{$^1$Department of Physics, North Dakota State University, Fargo, North Dakota 58108, United States}
\address{$^2$Center for Computational Materials Science, Naval Research Laboratory, Washington, D.C. 20375, United States}
\vspace{0.2cm}
\address{$^\ast$Corresponding author. E-mail: \textcolor[rgb]{0,0,1}{\mailto{khang.hoang@ndsu.edu}}}

\begin{abstract} 

Understanding the physics of structurally and chemically complex transition-metal oxide and polyanionic materials such as those used for battery electrodes is challenging, even at the level of pristine compounds. Yet these materials are also prone to and their properties and performance are strongly affected or even determined by crystallogra-phic point defects. In this review, we highlight recent advances in the study of defects and doping in such materials using first-principles calculations. The emphasis is on describing a theoretical and computational approach that has the ability to predict defect landscapes under various synthesis conditions, provide guidelines for defect characterization and defect-controlled synthesis, uncover the mechanisms for electronic and ionic conduction and electrochemical extraction and (re-)insertion, and provide an understanding of the effects of doping. Though applied to battery materials here, the approach is general and applicable to any materials in which the defect physics plays a role or drives the properties of interest. Thus, this work is intended as an in-depth review of defect physics in particular classes of materials, but also as a methodological template for the understanding and design of complex functional materials. 

\end{abstract}

\vspace{0.3pc}

\hspace*{1.9cm} compounds, electronic and ionic transport, delithiation mechanism 
%polarons, first-principles calculations

%\submitto{\JPCM}

%\maketitle


\vspace{-0.5pc}

\tableofcontents

\section{Introduction}

Materials for energy storage and conversion applications such as lithium-ion and sodium-ion batteries \cite{Whittingham_CR,Ellis2010,Masquelier2013CR,Palomares2013EES}, supercapacitors \cite{Zuo2017AS,Crosnier2018COE}, and solid-oxide fuel cells \cite{Kuklja2013PCCP,Kilner2014ARMR,Gao2016EES} are often complex transition-metal oxides and polyanionic compounds. As with other transition-metal oxides, they possess a rich physics arising from a subtle interplay among charge, spin, and lattice degrees of freedom \cite{Rao1989ARPC,Rabe2010ARCMP}. These complex materials are also known to be prone to and their properties are strongly affected by crystallographic point defects. In lithium-ion batteries, for example, the presence of some of these defects in the electrode materials can be vital or detrimental to the battery performance \cite{Schoonman199944,Ellis2006,Maxisch:2006p103,maier2008,Axmann2009CM,Hoang2011CM}. A detailed understanding of defect thermodynamics and kinetics, including the synthesis--(defect) structure--property relationship, is thus essential to explaining, predicting, and optimizing the materials' properties, and ultimately to the design and discovery of new materials with better performance. Progress in this area would also provide a better understanding of defect physics in strongly correlated functional materials in general. 

Over the past two decades, computational studies have been indispensable in the development of a fundamental understanding of the physics and chemistry of battery materials. First-principles calculations based on density-functional theory (DFT) \cite{HK,KS} have produced important results for and physical insights into various aspects of the materials, including atomic and electronic structure, voltage, capacity and cycling stability, power and rate capability, and thermal stability and safety \cite{Meng2009EES,Islam2014CSR,Urban2016npj}. These studies focused not only on bulk properties but also on the migration of alkali ions (often in conjunction with lithium vacancies) and, in some cases, small hole polarons and the effects of particular defects (e.g., antisite pairs) on ionic diffusion \cite{VanderVen2001,Morgan2004,Maxisch:2006p103,Kang2006,Ong2011PRB,Johannes2012PRB,Malik2010}. Far fewer studies have addressed defect physics and chemistry in a comprehensive manner.

Prior to 2011, several extensive studies of defects and doping in battery electrode materials were reported \cite{Islam:2005p168,Fisher:2008p80,Gardiner2010CM}. They were based on the traditional Kr\"oger-Vink, defect-reaction approach \cite{KV} and interatomic-potential simulations. The first systematic studies based on first-principles defect calculations were carried out for LiFePO$_4$ in 2011 \cite{Hoang2011CM,Hoang2012JPS}. Results from this work helped reconcile conflicting experimental reports on defect formation in the material, predicted the conditions under which LiFePO$_4$ with the lowest concentration of the detrimental iron antisite defects can be obtained, uncovered the electronic and ionic conduction mechanisms, and provided an understanding of the effects of doping. More importantly, the work demonstrated how systematic first-principles defect calculations can provide a comprehensive understanding of complex battery materials. Since 2012, there have been reports of extensive first-principles defect studies of a number of different materials, from those for lithium-ion battery cathodes \cite{Koyama2012,Koyama2013,Hoang2014JMCA,Koyama2014,Park2014,Santana2014,HoangLiMn2O4,Hoang2015PRA,Kong2015JPCC,Kong2015JMCA,Hoang2016CM,Hoang2017PRMoxides,Hoang2017Li2MnO3} and anodes \cite{Duan2015JPCC,Duan2015CMS,Cho2017JPCC} to lithium-air \cite{Radin2013EES,Varley2014EES,Radin2015CM}. Specific computational methods used in these studies vary from DFT within the local-density (LDA) or generalized gradient (GGA) approximation \cite{LDA1980,PW91} and the DFT+$U$ extension \cite{anisimov1991} to a hybrid DFT/Hartree-Fock approach \cite{Perdew1996JCP}.

This article reviews recent progress in the study of defects and doping in lithium-ion battery electrode materials, emphasizing the use of {\it defect physics as a theoretical framework for understanding and designing complex functional materials} in general. In this approach, state-of-the-art and systematic first-principles defect calculations can be regarded as well-controlled computational experiments carried out to probe complex materials at the electronic and atomic level. Section \ref{sec;approach} will outline the computational approach for defect studies. The power of the theoretical and computational approach will be illustrated in sections \ref{sec;electronic}--\ref{sec;doping} with select examples drawn mainly from our work on layered transition-metal oxides, olivine phosphates, and spinel-type oxides. Direct comparisons with experiments will also be emphasized, particularly regarding defect characterization, defect-controlled synthesis, and electronic and ionic conductivities. Finally, we will end this topical review with conclusions and outlook in section \ref{sec;conclusions}.

\section{Computational approach}\label{sec;approach}

A study of defect physics in a solid necessarily begins with an investigation of the {\it host} compound which acts as a reference for defect calculations (Note that the host material does {\it not} have to be a perfect stoichiometric compound). This includes the calculation of the atomic and electronic structure, phase stability, and any other bulk properties that may be deemed necessary to understand the physics of the host material. In subsequent calculations, defects\footnote{We often use the word ``defect'' as a generic term, referring to not only native defects intrinsic to the materials but also impurities (extrinsic defects) and defect complexes. Impurities, not to be confused with ``impurity phases,'' can be intentionally added (i.e., dopants) or unintentionally present.} are treated within supercell models, in which a defect is included in a periodically repeated finite volume of the host compound which itself contains many original unit cells. The total-energy electronic structure calculations are based on DFT, often using the DFT+$U$ extension \cite{anisimov1991} or a hybrid DFT/Hartree-Fock approach \cite{Perdew1996JCP} since it is known that DFT within LDA/GGA \cite{LDA1980,PW91} fails catastrophically in localized electron systems and particularly for transition-metal oxides. The hybrid functional approach is usually the method of choice because it treats {\it all} orbitals in the material on equal footing, unlike DFT$+U$ where {\it a priori} knowledge of parameters like the ``Hubbard $U$'' for each orbital in each element in each local chemical environment is necessary. There is ample evidence that the electronic structure of complex materials is better represented by hybrid functionals than by other DFT approaches \cite{Johannes2012PRB,Franchini2014JPCM}. Unless otherwise noted, the work we present in this review employs a specific implementation known as the HSE06 screened hybrid functional \cite{heyd:8207} available in the \textsc{vasp} code \cite{vasp2}.

The {\it formation energy} of a general defect X in charge state $q$ (with respect to the host lattice) is defined as \cite{Zhang1991PRL,VdW1993PRB,walle:3851,Freysoldt2014RMP}
\begin{equation}\label{eq;eform}
E^f({\rm X}^q)=E_{\rm tot}({\rm X}^q)-E_{\rm tot}({\rm host})-\sum_{i}{n_i\mu_i^\ast}+q(E_{\rm v}+\mu_{e})+ \Delta^q ,
\end{equation}
where $E_{\rm tot}({\rm X}^q)$ and $E_{\rm tot}({\rm host})$ are the total energies of a supercell containing the defect and the defect-free supercell, respectively. $\mu_i^\ast$ is the atomic chemical potential, accounting for the species $i$ either added ($n_i>0$) or removed ($n_i<0$) from the supercell to form the defect and representing the chemical reservoir with which the species is exchanged. $\mu_e$ is the electronic chemical potential, i.e., the Fermi level, which is the energy of the reservoir for electron exchange. As a convention, the Fermi level is referenced to the valence-band maximum (VBM) of the host ($E_{\rm v}$). The chemical potentials can be regarded as variables; however, they are not free parameters. In fact, $\mu_i^\ast$ is subject to thermodynamic constraints and can be used to represent experimental conditions, e.g., during preparation or use of the material; $\mu_e$ is determined by the charge neutrality condition that involves all positively and negatively charged defects and free electrons and holes, if present, in the material (and can then be compared with the Fermi-level position obtained in experiments; see, e.g., \cite{Doux2018ACSAEM}). Finally, $\Delta^q$ is a correction term to align the electrostatic potentials of the defect-free and defect supercells and to account for finite-size effects on the total energies of charged defects \cite{Freysoldt2009}.

In thermodynamic equilibrium, the formation energy of a defect directly determines the concentration \cite{walle:3851}:
\begin{equation}\label{eq;con} 
c=N_{\rm sites}N_{\rm config}\exp{\left(\frac{-E^f}{k_{\rm B}T}\right)}, 
\end{equation} 
where $N_{\rm sites}$ is the number of high-symmetry sites in the lattice (per unit volume) on which the defect can be incorporated, $N_{\rm config}$ is the number of equivalent configurations (per site), and $k_{\rm B}$ is the Boltzmann constant. Clearly, at a given temperature, a defect that has a lower formation energy will occur with a higher concentration. Note that the energy in equation (\ref{eq;con}) is, in principle, a free energy. However, the entropy and volume terms are neglected because they are usually small in solid phases. Besides, there is often significant cancellation between those terms in the host and in the reservoir \cite{walle:3851,Freysoldt2014RMP}. 

\begin{figure}[htb!]
\centering
\includegraphics[width=9.0cm]{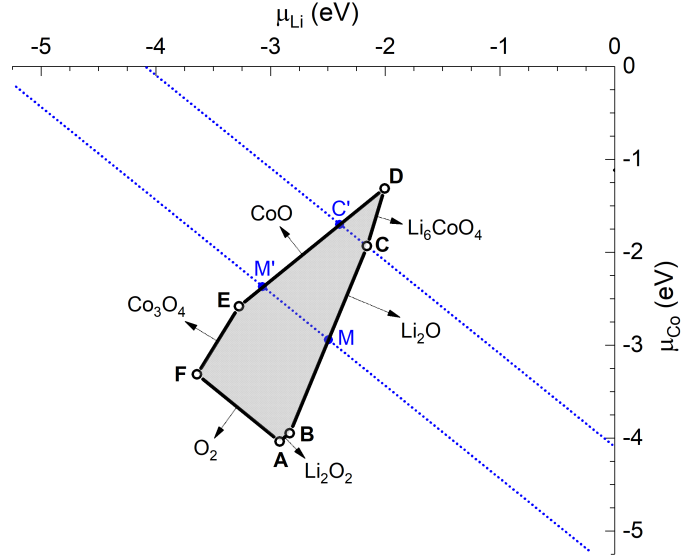}
\vspace{-0.1cm}
\caption{Chemical-potential phase diagram for LiCoO$_2$. Only Li--Co--O phases that define the stability region of LiCoO$_2$, shown as a shaded polygon, are indicated. The dotted (blue) lines correspond to the $\mu_{\rm O}$ levels associated with oxygen in air at about 400$^\circ$C (along $MM'$) and 900$^\circ$C ($CC'$). Produced with data from \cite{Hoang2014JMCA}.}\label{lco;pd}
\end{figure}

Let us illustrate with a specific example, namely a lithium vacancy in charge state $q$ in the battery material LiCoO$_2$:
\begin{equation}\label{eq;vli}
E^f(V_{\rm Li}^q)=E_{\rm tot}(V_{\rm Li}^q)-E_{\rm tot}({\rm host})+\mu_{\rm Li}^\ast+q(E_{\mathrm{v}}+\mu_{e})+ \Delta^q .
\end{equation}
Here, $\mu_{\rm Li}^\ast = E_{\rm tot}({\rm Li}) + \mu_{\rm Li}$, with $E_{\rm tot}({\rm Li})$ being the total energy per atom of metallic Li. All the quantities in equation (\ref{eq;vli}) can be obtained directly from the total-energy calculations, except $\mu_{\rm Li}$ and $\mu_e$. Although $\mu_{\rm Li}$ cannot be calculated directly without making specific assumptions about the lithium reservoir, the upper and lower bounds on its values, as well as those for $\mu_{\rm Co}$ and $\mu_{\rm O}$, can be determined by requiring that the host compound LiCoO$_2$ is stable against competing Li--Co--O phases \cite{Hoang2014JMCA}. Figure \ref{lco;pd} shows the allowed range of the $\mu_{\rm Li}$ and $\mu_{\rm Co}$ values, bound in a polygon in the two-dimensional ($\mu_{\rm Li}$, $\mu_{\rm Co}$) space. For a given point in the polygon, the remaining $\mu_{\rm O}$ variable is determined via the stability condition for the host \cite{Hoang2014JMCA}: 
\begin{equation}\label{eq;lco;stability}
\mu_{\rm Li} + \mu_{\rm Co} + 2\mu_{\rm O} = \Delta H({\rm LiCoO}_2), 
\end{equation}
where $\Delta H$ is the formation enthalpy calculated from DFT-based total energies. The determination of the Fermi level for a given set of $\mu_i$ values corresponding to a given point in the stability region will be made clear in section \ref{sec;layered} when we discuss the energetics of all relevant defects in the material.

It should be noted that a phase diagram such as figure \ref{lco;pd} is constructed based on zero-temperature energies. This, however, does not mean that temperature effects on the formation energy are all ignored. In fact, they can be included in the chemical potentials. For example, $\mu_{\rm O}$ is related to temperature and pressure via the expression \cite{Reuter2001}
\begin{equation}\label{eq;muO} 
\mu_{\mathrm{O}}(T,p)=\mu_{\mathrm{O}}(T,p_{\circ}) + \frac{1}{2}k_{\rm B}T {\rm ln}\frac{p}{p_{\circ}}, 
\end{equation} 
where $p$ and $p_{\circ}$ are, respectively, the partial pressure and reference partial pressure of O$_2$ gas. The reference state of $\mu_{\rm O}(T,p)$ can be chosen to be half of the total energy of an isolated O$_2$ molecule at 0 K. In figure \ref{lco;pd}, $\mu_{\rm O} = 0$ eV along the $AF$ line, corresponding to the value at 0 K. More experimentally relevant values can be estimated by using the actual synthesis conditions which place even stronger bounds on the atomic chemical potentials. For example, LiCoO$_2$ is often synthesized at about 400--900$^\circ$C in air \cite{Gummow1992327}. This translates into a range of $\mu_{\rm O}$ values from $-0.74$ eV (along the $MM'$ line in figure \ref{lco;pd}) to $-1.40$ eV (the $CC'$ line) with the upper and lower bounds obtained from the Gibbs free energy of O$_2$ gas at 0.21 atm and 400$^\circ$C and 900$^\circ$C, respectively \cite{stull1971}. The stability region enclosed by points $M$, $C$, $C'$, and $M'$ in figure \ref{lco;pd} can be considered as representing approximately the actual conditions under which LiCoO$_2$ is prepared. Here, as a first approximation, temperature-dependent effects are considered only for the gaseous (O$_2$) phase; the effects for the solid phases are ignored as they are often small \cite{Freysoldt2014RMP}.

In the methodology just described, native defects, including {\it charged} defects, and their energetics can thus be studied {\it individually} and the dependence of their formation energy and hence concentration on the Fermi level and the relative abundance of the host compound's constituent elements in the synthesis environment can be explicitly examined. Effects of impurities, either intentionally added or unintentionally present in the material, can be examined in the same manner; see also section \ref{sec;doping}. It should be noted that, though first-principles defect calculations are often carried out using large supercell models and assuming the dilute defect limit, direct defect--defect interaction can be studied by explicitly considering defect complexes in the calculations. Also note that this approach for defect calculations is different from the traditional Kr\"oger-Vink approach where calculations are carried out only for neutral defects or neutral defect complexes, e.g., Frenkel and Schottky pairs, based on pre-selected defect reactions \cite{KV}. Any specific defect reaction, if thermodynamically possible, is already included in the chemical-potential phase diagram and corresponds to a point in the stability region.

To understand defect physics in complex energy materials and its implications on the materials' properties and performance, one needs to carry out systematic calculations for all possible electronic and ionic defects (polarons, vacancies, interstitials, and antisites) as well as impurities, if present. The calculations produce information about the atomic and electronic structure and energetics of each defect, and the {\it defect landscape} \cite{Hoang2011CM}, which shows the relative formation energy (and hence concentration) of all defects, as a function of the atomic chemical potentials, i.e., synthesis conditions. Migration barriers of defects, especially those that can participate in charge and mass transport, can be computed using standard techniques such as the nudged elastic band method \cite{ci-neb}. All this information from the calculations can then be employed to explore all possible defect landscapes, provide guidelines for defect characterization and defect-controlled synthesis, uncover the mechanisms for electronic and ionic conduction and electrochemical extraction and (re-)insertion, and understand the effects of doping.

\section{Electronic structure {\it vis-\`{a}-vis} polaron formation}\label{sec;electronic}

\begin{figure}[htb!]
\centering
\includegraphics[width=9.5cm]{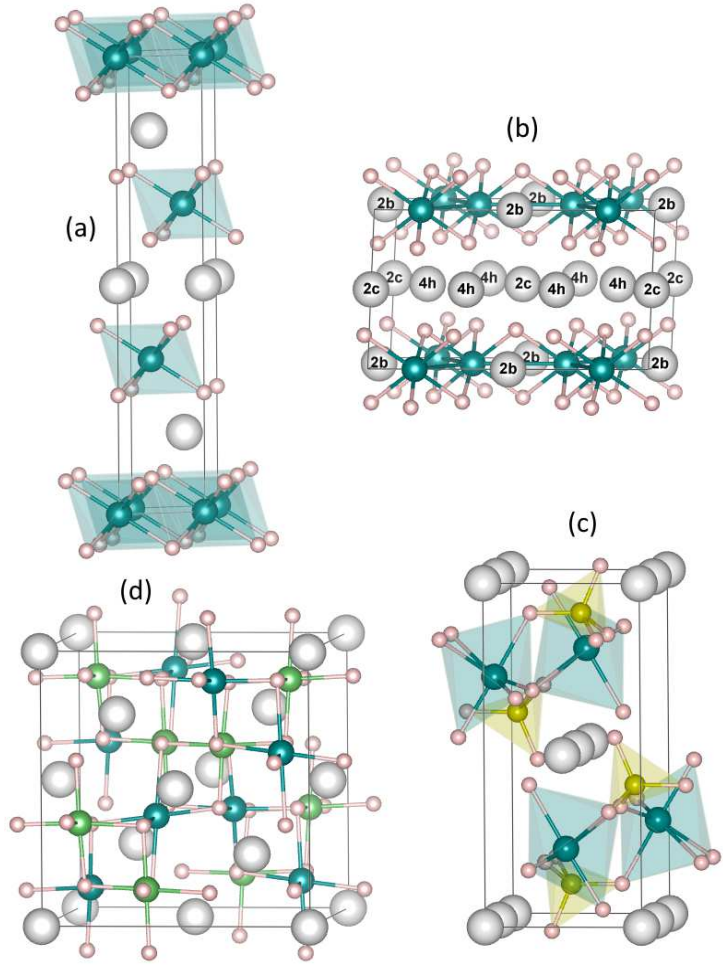}
\vspace{-0.1cm}
\caption{Atomic structure of (a) LiCoO$_2$ (trigonal, $R\bar{3}m$), (b) Li$_2$MnO$_3$ (monoclinic, $C2/m$), (c) LiFePO$_4$ (orthorhombic, $Pnma$), and (d) LiMn$_2$O$_4$ (slightly tetragonally distorted from cubic $Fd\bar{3}m$ \cite{HoangLiMn2O4}). Large (gray) spheres are Li, medium (blue or green) spheres are transition metals, small (yellow) spheres are P, and smallest (pink) spheres are O. The inequivalent Li Wyckoff positions in Li$_2$MnO$_3$ are indicated. In LiMn$_2$O$_4$, the Mn$^{3+}$ and Mn$^{4+}$ ions are presented as dark (blue) and light (green) medium spheres. All the structures are visualized using the \textsc{vesta} package \cite{VESTA}.}\label{struct}
\end{figure}

Before discussing defect physics in complex materials, let us first examine their atomic and electronic structure. Figure \ref{struct} shows the crystal structure of representatives from different classes of battery electrode materials: layered oxides, olivine phosphates, and spinel-type oxides. LiCoO$_2$ has alternate layers of Li$^+$ and (CoO$_2$)$^-$ in which cobalt is stable as low-spin Co$^{3+}$ and locates at the center of octahedra formed by oxygen. In a simple ionic model, the material can be regarded as an ordered arrangement of Li$^+$, Co$^{3+}$, and O$^{2-}$ \cite{Hoang2014JMCA}. The structure of LiNiO$_2$ and LiMnO$_2$ is slightly distorted compared to that of LiCoO$_2$, due to the strong Jahn-Teller effects associated with low-spin Ni$^{3+}$ and high-spin Mn$^{3+}$ ions \cite{Hoang2014JMCA,Hoang2015PRA}. In layered mixed transition-metal oxides LiNi$_{1/3}$Co$_{1/3}$Mn$_{1/3}$O$_2$ (NCM$_{1/3}$, also known as NMC$333$ or NMC$111$) and LiNi$_{1/3}$Co$_{1/3}$Al$_{1/3}$O$_2$ (NCA$_{1/3}$), not shown in the figure, the transition-metal layer contains a mixture of Ni$^{2+}$, low-spin Co$^{3+}$, and Mn$^{4+}$ (NCM$_{1/3}$) or low-spin Ni$^{3+}$, low-spin Co$^{3+}$, and Al$^{3+}$ (NCA$_{1/3}$) \cite{Hoang2016CM}. Li-rich Li$_2$MnO$_3$, also known as Li[Li$_{1/3}$Mn$_{2/3}$]O$_2$, has a layered structure similar to that of LiMnO$_2$ but with one-third of the atoms in the Mn layer replaced by lithium; as a result, each oxygen in Li$_2$MnO$_3$ has only two Mn neighbors, instead of three as in LiMnO$_2$. Manganese is stable as Mn$^{4+}$ and the material can be regarded as consisting of Li$^+$, Mn$^{4+}$, and O$^{2-}$ \cite{Hoang2015PRA}. In LiFePO$_4$, lithium forms one-dimensional channels and the transition metal locates at the center of a slightly distorted FeO$_6$ octahedron. Iron is stable as high-spin Fe$^{2+}$ and the compound can be regarded as consisting of Li$^+$, Fe$^{2+}$, and (PO$_{4}$)$^{3-}$ \cite{Hoang2011CM}. Finally, LiMn$_2$O$_4$ is a mixed-valence compound containing both Jahn-Teller active (high-spin Mn$^{3+}$) and non-active (Mn$^{4+}$) ions. This spinel-type oxide is known to possess a cubic structure at room temperature but transforms into a partially charge-ordered orthorhombic or tetragonal phase at low temperatures \cite{Yamada1995,Carvajal1998,Oikawa1998,Wills1999}. The supercell model with the Mn$^{3+}$/Mn$^{4+}$ ordering shown in figure \ref{struct}(d) was found to have the lowest total energy (at 0 K) among possible Mn$^{3+}$/Mn$^{4+}$ arrangements \cite{HoangLiMn2O4}.  

\begin{figure}[htb!]
\centering
\includegraphics[width=15.8cm]{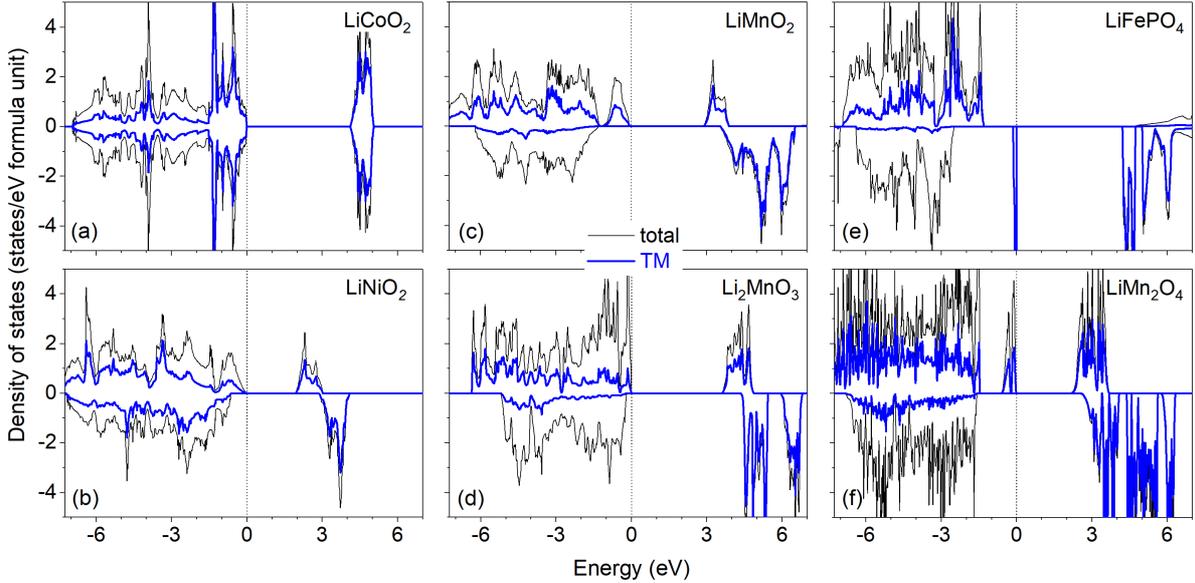}
\vspace{-0.6cm}
\caption{Total and partial densities of states of (a) LiCoO$_2$, (b) LiNiO$_2$, (c) LiMnO$_2$, (d) Li$_2$MnO$_3$, (e) LiFePO$_4$, and (f) LiMn$_2$O$_4$, obtained in HSE06 calculations. The ferromagnetic spin configuration was assumed for the transition-metal (TM) array in the lattice. The zero of the energy (at 0 eV) is set to the highest occupied state.}\label{dos}
\end{figure}    

In the vast majority of battery electrode materials, including those explicitly discussed in this review, the transition-metal ion, often also the redox center, is octahedrally coordinated with oxygen. This gives rise to the well-known crystal-field splitting of the five transition-metal $d$-states into an upper strong anti-bonding doublet ($e_g$) manifold with significant oxygen admixture, and a lower weak- or non-bonding ($t_{2g}$) triplet manifold with minimal oxygen admixture. These electronic states form both the valence and conduction bands and their dispersion and transition-metal vs.~oxygen content have a strong effect on the type of defects that can form, as well as on other battery properties such as voltage \cite{Hoang2016CM,Johannes2017Springer}, chemical stability \cite{Johannes2016SSI}, and transport efficiency \cite{Johannes2012PRB}. The nature of these bands is determined by the electron count (charge state) of the transition metal and by a competition between crystal-field splitting and Hund's rule which determines whether the system is in a high-spin or low-spin configuration. 

Figure \ref{dos} shows the electronic density of states (DOS) of the six compounds (including those shown in figure \ref{struct}). Both the total DOS and the partial DOS (PDOS) of the transition metal are plotted, with the two spin channels plotted separately on the positive and negative $y$-axis. The difference between the total DOS and the transition-metal PDOS can be understood as the amount of oxygen character since the Li states are high up in the conduction band. It is instructive to first compare two of the most common battery materials: LiCoO$_2$ and LiFePO$_4$. In the former, the Co ion is in the $+3$ charge state ($d^6$ configuration). The crystal field from the Co $3d$ and O $2p$ bonding is strong enough to overcome the Hund's coupling and produces a low-spin ground state ($t_{2g}^6e_g^0$) with all electrons in the lower weakly anti-bonding $t_{2g}$ complex. Due to the Co $d$--O $p$ interaction, this complex is fairly dispersive (broad DOS); see figure \ref{dos}(a). In LiFePO$_4$, the Fe ion is in the $+2$ state (also $d^6$ configuration). But, because the PO$_4$ unit is already tightly bound, the bonding between Fe and O is weak and the crystal-field splitting is small; as a result, the ground-state configuration is high spin ($t_{2g}^4e_g^2$). This leaves the highest energy electron in the minority spin anti-bonding $t_{2g}$ complex, like LiCoO$_2$, but with significantly less dispersion and oxygen admixture, giving rise to a sharply peaked Fe-dominant DOS just below the VBM; see figure \ref{dos}(e). 

As a general rule, the bonding between oxygen and transition metal increases as one moves right across the 3$d$ row of the periodic table. Within a materials (structural) class, then, a rightward shift of the transition-metal ion increases the crystal-field splitting and the amount of oxygen in the valence band, and decreases the tendency toward hole localization. For instance, in LiCoO$_2$, the Co character clearly dominates the DOS at the VBM. In LiNiO$_2$, by contrast, the Ni character is less than 50\% of the total \cite{Hoang2014JMCA}, reflecting the greater oxygen admixture; see figure \ref{dos}(b). LiMnO$_2$ here appears to violate this rule, showing rather strong oxygen character in the valence band despite being leftward of Co and Fe in the periodic table; see figure \ref{dos}(c). This is, however, precisely {\it because} of the weaker bonding between Mn $3d$ and O $2p$ states that weakens the crystal-field splitting, resulting in a high-spin rather than low-spin ground state. Electrons at the top of the valence band are therefore in the comparatively oxygen-rich $e_g$ states rather than the oxygen-poor $t_{2g}$. Regarding the other Mn-based compounds, the top of the valence band of LiMn$_2$O$_4$ is predominantly composed of the Mn $3d$ states, specifically from the Mn$^{3+}$ ions \cite{HoangLiMn2O4}, whereas that of Li$_2$MnO$_3$ is predominantly O $2p$ states \cite{Hoang2015PRA}. The much higher oxygen content in the case of Li$_2$MnO$_3$ also comes from the fact that oxygen in this oxide is undercoordinated, compared to that in LiMnO$_2$. For all compounds presented in figure \ref{dos}, the conduction-band bottom is predominantly composed of the transition-metal $3d$ states. The conduction-band bottom of LiMn$_2$O$_4$, in particular, consists predominantly of the Mn $3d$ states from the Mn$^{4+}$ ions \cite{HoangLiMn2O4}. For the electronic structure of mixed transition-metal oxides NCM$_{1/3}$ and NCA$_{1/3}$, see, e.g., \cite{Hoang2016CM}.    

In the discussion above, we paid particular attention to the electronic states near the band edges. This is because they are relevant to transport and electrochemical processes. For example, in an oxidation reaction, electrons are removed from the material which results in electron holes being introduced at the VBM; a reduction reaction, on the other hand, involves addition of electrons to the material and these electrons then reside at the conduction-band minimum (CBM). The type of electronic defects that can form depends on whether the holes (electrons) can localize into polarons or delocalize into the valence (conduction) band. The {\it necessary condition} for a hole (electron) to localize on a particular transition metal is that the contribution of that transition metal to the electronic states at the VBM (CBM) of the host compound is larger than or equal to that from any other transition metal in the supercell and is larger than the contribution from any oxygen atom. Given that interplay between electronic structure and polaron formation, it becomes essential to reproduce correctly the electronic structure.  

\begin{figure}[htb!]
\centering
\includegraphics[width=9.0cm]{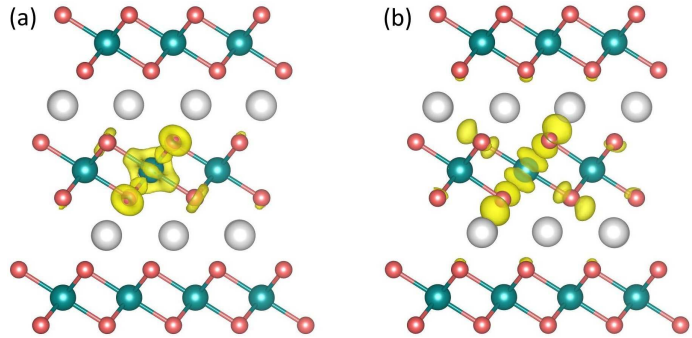}
%\vspace{-0.1cm}
\caption{Charge densities associated with hole and electron polarons in LiCoO$_2$ \cite{Hoang2014JMCA}: (a) $\eta^+$ and (b) $\eta^-$, corresponding to Co$^{4+}$ and Co$^{2+}$ at a Co$^{3+}$ lattice site. Large (gray) spheres are Li, medium (blue) spheres are Co, and smallest (red) spheres are O.}\label{lco;pola}
\end{figure}

Figure \ref{lco;pola} shows the local lattice environments and charge densities associated with hole ($\eta^+$) and electron ($\eta^-$) polarons\footnote{In this review, a hole (electron) polaron formed on a transition-metal ion is referred commonly to as $\eta^+$ ($\eta^-$); their true identities should be evident in the context of a specific host compound. Also note that, in the literature, hole and electron polarons are sometimes denoted as $p^+$ and $p^-$, respectively.} in LiCoO$_2$ \cite{Hoang2014JMCA}. $\eta^+$ and $\eta^-$ here are basically low-spin Co$^{4+}$ and high-spin Co$^{2+}$, respectively, at a Co$^{3+}$ site in the host lattice. These electronic defects are called free or {\it unbound} polarons as they are stable even in the absence of other native defects or impurities; they are also called {\it small polarons} \cite{Shluger1993} as the local lattice distortion induced by the presence of the localized hole or electron is limited mainly to the neighboring O atoms \cite{Hoang2014JMCA}. Unbound small polarons were also reported to occur in LiNiO$_{2}$ \cite{Hoang2014JMCA}, LiMnO$_2$ \cite{Hoang2015PRA}, NCM$_{1/3}$, NCA$_{1/3}$ \cite{Hoang2016CM}, LiFePO$_4$ \cite{Maxisch:2006p103,Hoang2011CM,Ong2011PRB}, LiMn$_2$O$_4$ \cite{HoangLiMn2O4}, consistent with the electronic structure discussed earlier, as well as in many other battery electrode materials. In Li-rich layered Li$_2$MnO$_3$, the addition of an electron also results in a small electron polaron ($\eta^-$), which is basically high-spin Mn$^{3+}$ at a Mn$^{4+}$ site. The removal of an electron results in a delocalized hole being introduced at the VBM, consistent with the fact the VBM of Li$_2$MnO$_3$ is predominantly composed of the relatively delocalized O $2p$ states. More interestingly, it is also because of that feature in the electronic structure that {\it bound} hole polarons ($\eta_{\rm O}^+$) at oxygen sites were found to form in the presence of other defects, e.g., negatively charged lithium vacancies ($V_{\rm Li}^-$) \cite{Hoang2015PRA}. This defect, first reported in \cite{Hoang2015PRA}, is essentially O$^-$ (with a calculated magnetic moment of 0.69$\mu_{\rm B}$) at an O$^{2-}$ site in the host lattice and can be referred to as ``bound oxygen hole polaron'' \cite{Hoang2015PRA}, ``localized electron hole states on oxygen'' \cite{Luo2016JACS,Luo2016NC}, or ``O$^-$ bound polaron'' \cite{Schirmer2006}. The formation of $\eta_{\rm O}^+$ provides a clear evidence for the anionic redox activity in Li$_2$MnO$_3$ and related materials; see sections \ref{sec;delithiation} and \ref{doping;deli} for further discussions of $\eta_{\rm O}^+$ and its implications on the delithiation mechanism and electronic conduction.

In this section, we have thus showed how one can tell from the electronic structure of a material if and where a hole or electron polaron can form. Whether the polaron can actually occur in the material with a significant concentration, when thermally activated, depends on its formation energy. A hole (electron) polaron is a charged defect, i.e., its formation energy is dependent on the position of the Fermi level [see equation (\ref{eq;eform})]. The determination of the polaron formation energy would thus require information about the energetics of all other possible defects that may occur in the material; see section \ref{sec;defects}. Polarons can also be activated during oxidation and/or reduction processes \cite{Hoang2015PRA}. Lithium extraction from a lithium-ion battery cathode during charging, for instance, often leads to simultaneous formation of hole polarons and negatively charged lithium vacancies in the electrode material; here, the more relevant quantity is the extraction voltage, instead of the defect formation energy; see section \ref{sec;delithiation}. The role of polarons as charge-carrying defects in (thermally activated) electronic conduction in complex materials will be discussed in section \ref{sec;conduction}. The interplay between electronic structure and polaron formation (and hence presence of active redox centers) has also been shown to be essential to the understanding of materials for pseudocapacitors \cite{Hoang2017PRM,Hoang2018RSC} and solid-oxide fuel cells \cite{Doux2018ACSAEM}.

\section{Defect energetics and tuning defect landscapes}\label{sec;defects}

In addition to the small hole and electron polarons discussed in section \ref{sec;electronic}, other possible native defects include vacancies, interstitials, and antisites. Defects can be {\it thermally activated}, e.g., during materials preparation at high temperatures and get trapped inside the materials when cooling down to room temperature. Certain defects may thus be present with significant concentrations in the materials and affect their properties and performance. Here, we discuss structure and energetics of native defects and the dependence of defect landscapes on the synthesis conditions, based on recent work on layered oxide \cite{Hoang2014JMCA,Hoang2015PRA,Hoang2016CM}, olivine phosphate \cite{Hoang2011CM}, and spinel-type oxide \cite{HoangLiMn2O4} materials.

\subsection{Layered oxides}\label{sec;layered}

Defects are common in layered oxide materials. LiCoO$_2$ prepared at low temperatures, for example, exhibits significant Co/Li disorder and poor electrochemical performance \cite{Gummow1992327}. Also, experimental studies of the magnetic properties always find the presence of localized magnetic moments in LiCoO$_2$ samples \cite{Chernova2011}. It has been reported that truly stoichiometric LiCoO$_2$ is possible; however, it requires rather delicate experimental procedures to prepare \cite{Menetrier2008ESSL}. In fact, the commercially available LiCoO$_2$ is often made deliberately Li-overstoichiometric. An even more striking example is that of LiNiO$_2$, for which the stoichiometric compound with all Ni$^{3+}$ does not really exist. The long-range Jahn-Teller distortion and magnetic ordering (associated with low-spin Ni$^{3+}$) expected in defect-free LiNiO$_2$ is absent in real samples and the material always has a significant concentration of Ni ions at the Li site \cite{Dutta1992123,Kanno1994216,Hirano1995,Barra1999,Chappel2002,Kalyani2005}. LiMnO$_2$, on the other hand, is prepared via ion exchange from NaMnO$_2$ because the synthesis at high temperatures often results in an orthorhombic phase \cite{Armstrong1996,Capitaine1996}. The material is known to have strong cation mixing \cite{Armstrong1996} and poor electrochemical performance which is usually ascribed to structural phase instabilities \cite{Vitins1997}. Manganese antisites have also been observed in Li-rich layered oxide Li$_2$MnO$_3$, i.e., Li[Li$_{1/3}$Mn$_{2/3}$]O$_2$, under certain synthesis conditions \cite{Kubota2012}. 

\begin{figure}[htb!]
\centering
\includegraphics[width=9.0cm]{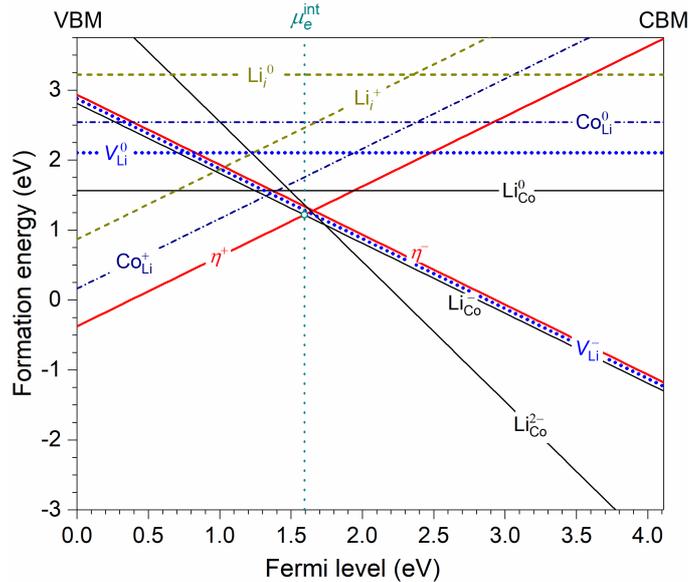}
\vspace{-0.1cm}
\caption{Formation energies of relevant native defects in LiCoO$_2$, plotted as a function of the Fermi level. The energies are obtained at point $M$ in the chemical-potential phase diagram (figure \ref{lco;pd}). The slope indicates the charge state [i.e., $q$ in equation (\ref{eq;eform})]: positively (negatively) charged defects have positive (negative) slopes. $\mu_e^{\rm int}$ is the position of the Fermi level determined by the charge neutrality condition involving the native/intrinsic defects. Produced with data from \cite{Hoang2014JMCA}.}\label{lco;fe}
\end{figure} 

Figure \ref{lco;fe} shows the formation energies of relevant defects in LiCoO$_2$ reported in \cite{Hoang2014JMCA}. These defects include unbound small hole ($\eta^+$, i.e., low-spin Co$^{4+}$) and electron ($\eta^-$, i.e., high-spin Co$^{2+}$) polarons mentioned in section \ref{sec;electronic}, lithium vacancies ($V_{\rm Li}$) and interstitials (Li$_i$), and lithium (Li$_{\rm Co}$) and cobalt (Co$_{\rm Li}$) antisites; other defects are not included because they have much higher energies (at least in the bulk; more discussion later). Each defect type may have several charge states, corresponding to different values for $q$ in equation (\ref{eq;eform}). However, among the ionic defects included in figure \ref{lco;fe}, only the following defect configurations are the true charge states: $V_{\rm Li}^-$ (i.e., a void formed by the removal of a Li$^+$), Li$_i^+$ (additional Li$^+$ ion at an interstitial site), Li$_{\rm Co}^{2-}$ (Li$^+$ replacing Co$^{3+}$ at a Co site), and Co$_{\rm Li}^+$ (high-spin Co$^{2+}$ replacing Li$^+$ at a Li site). These configurations are regarded as {\it elementary} defects. Other charge states are, in fact, defect complexes consisting of the elementary ionic defects and polaron(s). For example, the neutral charge state of $V_{\rm Li}$, nominally denoted as $V_{\rm Li}^0$, is a complex of $V_{\rm Li}^-$ and $\eta^+$; similarly, Co$_{\rm Li}^0$ is a complex of Co$_{\rm Li}^+$ and $\eta^-$, and Li$_{\rm Co}^0$ is a complex of Li$_{\rm Co}^{2-}$ and two $\eta^+$ \cite{Hoang2014JMCA}.

In the absence of electrically active impurities or when they occur with lower concentrations than the charged native defects, the Fermi level of LiCoO$_2$ is at $\mu_e = \mu_e^{\rm int}$ (``int'' stands for ``intrinsic'') where charge neutrality is maintained, i.e., positive and negative charges originating from the native defects are balanced. Because of the strong (exponential) dependence of the concentration on the formation energy, $\mu_e^{\rm int}$ is essentially ``pinned'' at the position where the lowest-energy positively and negatively charged defects have equal formation energies. In the example presented in figure \ref{lco;fe}, $\mu_e^{\rm int}$ is predominantly determined by $\eta^+$ and Li$_{\rm Co}^-$. The energetically most stable configurations of the vacancies, interstitials, and antisites at $\mu_e^{\rm int}$ are $V_{\rm Li}^-$, Li$_i^+$, Li$_{\rm Co}^-$ (a complex of Li$_{\rm Co}^{2-}$ and $\eta^+$), and Co$_{\rm Li}^+$. Cobalt is thus most stable as high-spin Co$^{2+}$ at the Li site, as opposed to being low-spin Co$^{3+}$ at its original (Co) lattice site. It is noted that $\mu_e^{\rm int}$ is far from both the band edges; as a result, band-like carriers would be negligible. In fact, the Fermi level of the system cannot come close to the VBM or CBM, otherwise the formation energy of certain charged native defects will become very small or even negative, see figure \ref{lco;fe}, and the host compound will become unstable. This has important implications for the electronic conduction and doping mechanisms as discussed in sections \ref{sec;conduction} and \ref{sec;doping}.

The defect landscape shown in figure \ref{lco;fe} is not the only scenario that may occur. In fact, as reported in \cite{Hoang2014JMCA}, defect energetics is sensitive to the choice of the atomic chemical potentials and hence the synthesis conditions. Under the conditions at point $M'$ in figure \ref{lco;pd}, for example, the lowest-energy native defects in LiCoO$_2$ are Co$_{\rm Li}^+$ and $V_{\rm Li}^-$, whereas at points $C$ and $C'$ they are Co$_{\rm Li}^+$ and $\eta^-$; for more results at other points in the chemical-potential phase diagram, see \cite{Hoang2014JMCA}. The scenario reported by Gummow {\it et al.}~\cite{Gummow1992327}, where significant Co/Li disorder was observed experimentally, could be identified with that obtained under the conditions somewhere between $M$ and $M'$ in the chemical-potential diagram. The experimental preparation of Li-overstoichiometric LiCoO$_2$ is likely carried out under the conditions somewhere near $M$ and between points $M$ and $C$ where the host is in equilibrium with Li$_2$O as the impurity phase is often observed when preparing LiCoO$_2$ in Li-excess (Co-deficient) environments \cite{Levasseur2003,Menetrier2008ESSL}. From the defect landscape in figure \ref{lco;fe}, the chemical formula for the Li-overstoichiometric samples can be written as Li$_{1+\delta}$Co$_{1-\delta}$O$_{2}$ or, more explicitly, as Li[Co$_{1-3\delta}^{3+}$Li$_{\delta}^+$Co$_{2\delta}^{4+}$]O$_2$ where each Li$_{\rm Co}^{2-}$ is charge compensated by two $\eta^{+}$, assuming that other defects have negligible concentrations. The excess Li thus goes into the Co site, instead of an interstitial site. This is because Li interstitials are energetically less favorable in LiCoO$_2$ \cite{Hoang2014JMCA}, as also seen in figure \ref{lco;fe}. The results reviewed here thus indicate that {\it one can tune the defect landscape in the material by tuning the synthesis conditions}. Note that, as the defect landscape changes, the Fermi-level position $\mu_e^{\rm int}$ changes accordingly \cite{Hoang2014JMCA}. Also, in all defect landscapes, the lowest-energy native defects in the material always contain low-spin Co$^{4+}$ (in the form of the $\eta^+$ defect) and/or high-spin Co$^{2+}$ (in the form of $\eta^-$ or Co$_{\rm Li}^+$) \cite{Hoang2014JMCA}. This is consistent with the presence of localized magnetic moments observed in experiments \cite{Chernova2011}.    

\begin{figure}[htb!]
\centering
\includegraphics[width=9.0cm]{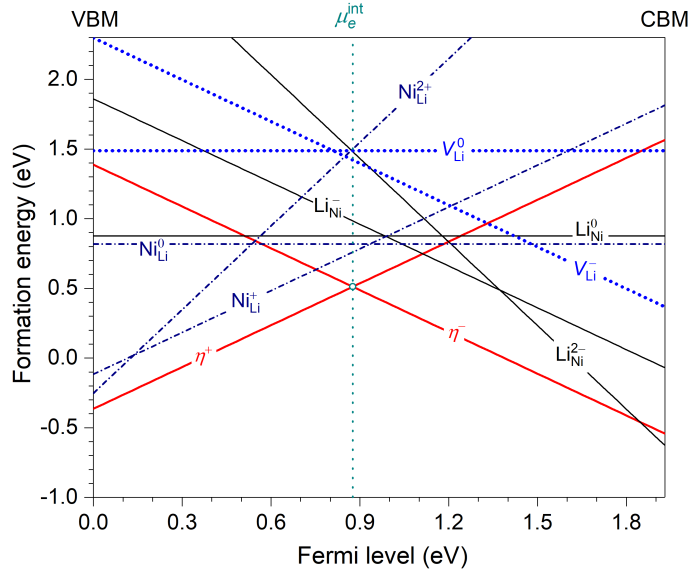}
\vspace{-0.1cm}
\caption{Formation energies of relevant native defects in LiNiO$_2$. The energies are obtained at a point in the chemical-potential phase diagram where LiNiO$_2$ is in equilibrium with Li$_2$O and Li$_2$NiO$_2$. Produced with data from \cite{Hoang2014JMCA}.}\label{lno;fe}
\end{figure} 

Defect landscapes in different materials are often very different. Figure \ref{lno;fe} shows a representative landscape in LiNiO$_2$. As reported in \cite{Hoang2014JMCA}, unbound small polarons $\eta^{+}$ (i.e., low-spin Ni$^{4+}$) and $\eta^{-}$ (i.e., Ni$^{2+}$) are always the lowest-energy native defects, independent of the atomic chemical potentials. These polarons ``pin'' the Fermi level of LiNiO$_2$ at $\mu_e^{\rm int}$, where charge neutrality is maintained, and have a formation energy of only 0.51 eV. $\eta^{+}$ and $\eta^{-}$ in LiNiO$_2$ can be activated via an {\it electron--hole polaron pair} mechanism (somewhat similar to the Frenkel pair mechanism for ionic defects \cite{amidePRB}); here, one electron is transferred from one Ni$^{3+}$ to another which results in the formation of Ni$^{4+}$ and Ni$^{2+}$, i.e., the $\eta^{+}$--$\eta^{-}$ pair. Our explicit calculations for this polaron pair, in which the two Ni sites are nearest neighbors, show that it has a formation energy of 0.46 eV. With such a low energy, $\eta^{+}$ and $\eta^{-}$ are easy to form, i.e., a certain amount of Ni$^{3+}$ ions, estimated to be about 8\% when the material is prepared at 800$^\circ$C, undergo {\it charge disproportionation}: 2Ni$^{3+}$ $\rightarrow$ Ni$^{4+}$ + Ni$^{2+}$. Note that in the concentration estimation, $N_{\rm config} = 6$ for the $\eta^{+}$--$\eta^{-}$ pair and thermodynamic equilibrium is assumed. The nickel antisite Ni$_{\rm Li}^{+}$ (i.e., Ni$^{2+}$ at a Li site) also has a low formation energy \cite{Hoang2014JMCA}. These results explain the experimental observations mentioned earlier, including the difficulty in synthesizing LiNiO$_2$ with all Ni$^{3+}$ at high temperatures, the absence of long-range Jahn-Teller distortion and magnetic ordering, and the presence of nickel antisites \cite{Dutta1992123,Hirano1995,Kanno1994216,Barra1999,Chappel2002,Kalyani2005}. Note that, although the formation energy of $\eta^{+}$--$\eta^{-}$ (as well as that of $\eta^{+}$ and  $\eta^{-}$ at $\mu_e^{\rm int}$) is independent of the atomic chemical potentials, i.e., one cannot eliminate the charge disproportionation by simply tuning the synthesis conditions \cite{Hoang2014JMCA}, the concentration of Ni$^{4+}$ and Ni$^{2+}$ may be reduced by lowering the synthesis temperature, see equation (\ref{eq;con}), assuming that the system is still in (or close to) thermodynamic equilibrium.

Regarding the Mn-based layered oxides, the lowest-energy positively and negatively charged native defects in LiMnO$_2$ can be the small hole polaron $\eta^{+}$ (i.e., Mn$^{4+}$) and $V_{\rm Li}^{-}$, $\eta^{+}$ and Li$_{\rm Mn}^{2-}$, Mn$_{\rm Li}^{+}$ and Li$_{\rm Mn}^{2-}$, or Mn$_{\rm Li}^{+}$ and $V_{\rm Li}^{-}$, depending on the chosen set of the atomic chemical potentials. Under all possible synthesis conditions, the defect landscape of the material is characterized by the presence of low-formation-energy antisites Mn$_{\rm Li}^{+}$ and Li$_{\rm Mn}^{2-}$ \cite{Hoang2015PRA}. These findings are thus consistent with experiments showing strong cation mixing in LiMnO$_2$ \cite{Armstrong1996,Capitaine1996}. As discussed in \cite{Hoang2015PRA}, the antisites can act as nucleation sites for the formation of orthorhombic LiMnO$_2$ during synthesis or spinel LiMn$_2$O$_4$ during electrochemical cycling, which leads to inferior cycling stability \cite{Vitins1997}. In Li$_2$MnO$_3$, Mn$_{\rm Li}^+$ (i.e., high-spin Mn$^{2+}$ at the Li site, $4h$ or $2c$) was also found to have a low formation energy under Li-deficient and/or reducing conditions \cite{Hoang2015PRA}, which is consistent with the presence of these antisites in the ``oxygen-deficient Li$_{2}$MnO$_{3-x}$'' samples synthesized in the presence of strong oxygen-reducing agents \cite{Kubota2012}. Most interestingly, $V_{\rm Li}^0$ in Li$_2$MnO$_3$ was found to be a complex of $V_{\rm Li}^-$ and $\eta_{\rm O}^+$ where the latter is a hole polaron stabilized on an oxygen and bound to the former \cite{Hoang2015PRA}. This is completely different from the structure of $V_{\rm Li}^0$ found in LiMO$_2$ (M = Co, Ni, Mn) and in the vast majority of electrode materials where the hole is localized on a transition-metal ion. The implications of this on the delithiation mechanism and electronic conduction are discussed in sections \ref{sec;delithiation} and \ref{doping;deli}.  

In layered mixed transition-metal oxides such as NCM$_{1/3}$ and NCA$_{1/3}$, antisites were also reported to be common \cite{Hoang2016CM}. It is believed, however, that a small concentration of transition-metal antisites can be benign or even beneficial through the so-called ``pillaring effect'' that enhances the structural stability of layered oxides \cite{Yoon2018CM,Xiao2008CM,Chernova2011,Liang2014}. The low formation energy of antisites in all the layered oxide materials discussed here can be ascribed in part to the small ionic-radius difference between the transition-metal ion (specifically, high-spin M$^{2+}$--the stable configuration of the transition-metal ions at the Li lattice site) and the Li$^+$ ion. However, it was also found that the formation energy has a strong dependence on the relative abundance of the host's constituent elements in the synthesis environment, i.e., the atomic chemical potentials \cite{Hoang2014JMCA,Hoang2015PRA,Hoang2016CM}, as also demonstrated in this section. Systematic and comprehensive studies of defects following the computational approach outlined in section \ref{sec;approach} are thus necessary to provide guidelines for defect characterization and defect-controlled synthesis of complex materials.

Note that, in the examples discussed in this review, the calculations were carried out for defects in the bulk. Certain defects such as transition-metal and oxygen vacancies were often found to be energetically unfavorable in the interior of the materials \cite{Hoang2014JMCA,Hoang2015PRA,Hoang2016CM}. This is because the creation of these defects involves breaking the strong transition-metal--oxygen covalent bonds which requires a high energy. At/near surfaces or interfaces, however, defects such as oxygen vacancies can have a much lower formation energy, given the less constrained lattice environment \cite{Hoang2014JMCA,Hoang2015PRA}. It is then important to determine if non-stoichiometry observed in a given material is due to defects at the surface/interface or in the bulk. Also, even defects with high formation energies can occur with high concentrations if the material is prepared using non-equilibrium methods which can lead to excess defects being frozen in. In this case, the equilibrium concentration estimated via equation (\ref{eq;con}) should only be regarded as the lower bound.     

\subsection{Olivine phosphates}

Iron antisites (Fe$_{\rm Li}$) are often observed in LiFePO$_4$. These relatively immobile defects can block the one-dimensional lithium channels and thus reduce the electrochemical performance \cite{Yang2002,chen2008,Chung2008,Axmann2009CM}. The antisites can occur simultaneously with other native defects. Chung {\it et al.}~\cite{Chung2008}, for example, reported evidence of a certain amount of iron and lithium atoms exchanging sites and forming antisite pairs Fe$_{\rm Li}$--Li$_{\rm Fe}$, while Axmann {\it et al.}~\cite{Axmann2009CM} found that Fe$_{\rm Li}$ is formed in association with lithium vacancies ($V_{\rm {Li}}$). These conflicting reports suggest that the results can be sensitive to the actual synthesis conditions. This is, indeed, the case, as shown in a first-principles study reported in \cite{Hoang2011CM} which explored all defect landscapes under different conditions. The work also provided guidelines for obtaining LiFePO$_4$ with the lowest possible concentration of the iron antisites, which was eventually confirmed by experiments \cite{Park2016EES}. Antisite defects have also been observed in other olivine phosphates such as LiCoPO$_4$ \cite{Allen2011JPS,Truong2013AMI,Boulineau2015CM} and LiNiPO$_4$ \cite{Devaraju2015SR,Biendicho2017IC}.

\begin{figure}[htb!]
\centering
\includegraphics[width=15.8cm]{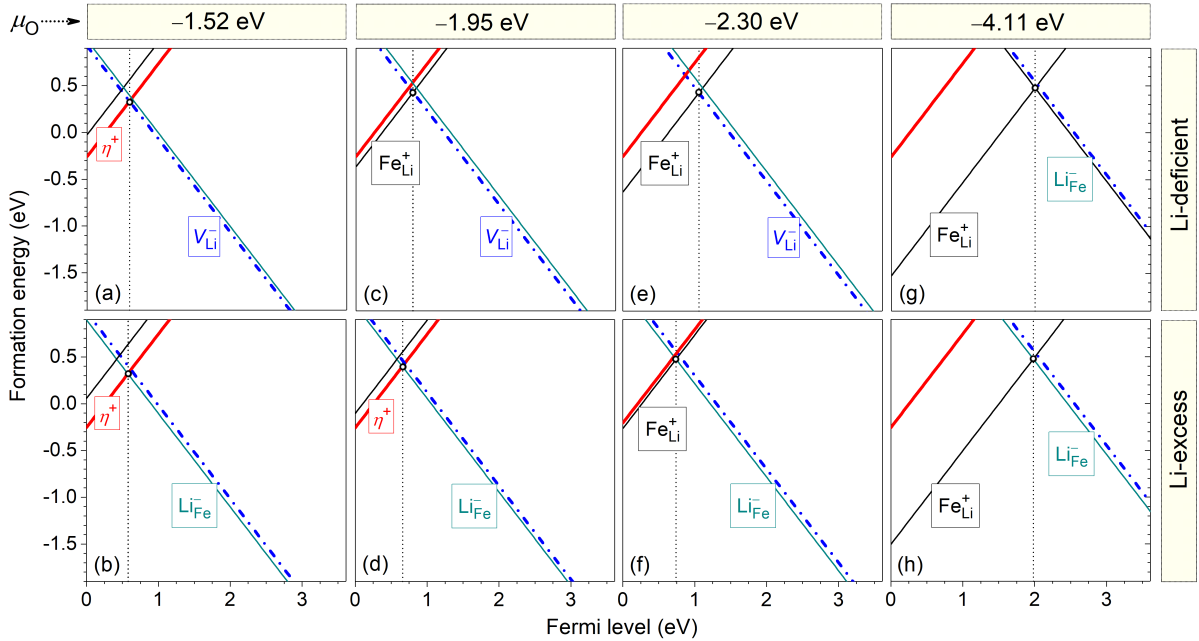}
\vspace{-0.6cm}
\caption{Formation energies of low-energy positively ($\eta^+$, Fe$_{\rm Li}^+$) and negatively ($V_{\rm Li}^-$, Li$_{\rm Fe}^-$) charged, elementary defects in LiFePO$_4$, plotted as a function of the Fermi level from the VBM to CBM. The energies are obtained under different synthesis conditions, i.e., different $\mu_{\rm O}$ values and in the Li-deficient or Li-excess environment. The vertical dotted line marks the Fermi-level position $\mu_{e}^{\rm{int}}$ of the material, determined predominantly by the lowest-energy charged defects. Produced with data from \cite{Hoang2011CM}.}\label{lfp;fe}
\end{figure}

Relevant elementary native defects in LiFePO$_4$ include the small hole polaron $\eta^+$ (i.e., high-spin Fe$^{3+}$ at a Fe$^{2+}$ lattice site; denoted as $p^+$ in \cite{Hoang2011CM} and \cite{Hoang2012JPS}), $V_{\rm Li}^-$, Li$_i^+$, Fe$_{\rm Li}^+$ (high-spin Fe$^{2+}$ at a Li site), and Li$_{\rm Fe}^-$ (Li$^+$ at a Fe site). Other charge states of these defects are complexes consisting of the elementary ionic defects and polaron(s), similar to defects in layered oxides. For example, $V_{\rm Li}^0$ is a complex of $V_{\rm Li}^-$ and $\eta^+$, whereas Li$_{\rm Fe}^0$ is a complex of Li$_{\rm Fe}^-$ and $\eta^+$ \cite{Hoang2011CM}. Different possible defect landscapes were explored by varying the atomic chemical potential values bound in slices of a polyhedron in the three-dimensional phase space in which LiFePO$_4$ is thermodynamically stable. One focuses on two ``knobs'' that can be used to tune the synthesis conditions in practice: one is the oxygen chemical potential $\mu_{\rm O}$, which can be controlled by controlling temperature and pressure and/or by using oxygen-reducing agents, and the other is the relative abundance of lithium in the environment, ranging from ``Li-deficient'' to ``Li-excess'' \cite{Hoang2011CM}.

Figure \ref{lfp;fe} shows different defect landscapes in LiFePO$_4$ associated with different sets of the atomic chemical potentials, ranging from the lowest to highest possible $\mu_{\rm O}$ value and from the Li-deficient to Li-excess environment, reported in \cite{Hoang2011CM}; for simplicity, only the lowest-energy charged defects are shown. The results were obtained in DFT+$U$ calculations. Lower $\mu_{\rm O}$ values (more reducing environments) are usually associated with higher temperatures and/or lower oxygen partial pressures [equation (\ref{eq;muO})] and/or the presence of oxygen-reducing agents; in the extreme Li-excess (Li-deficient) environment, the system is close to forming Li-containing (Fe-containing) impurity phases \cite{ong2008}. The results show that Fe$_{\rm Li}^+$ is the lowest-energy positively charged defect in a majority of the defect landscapes. The iron antisite can occur simultaneously with $V_{\rm Li}^-$ to form the neutral complex Fe$_{\rm Li}^+$--$V_{\rm Li}^-$ whose formation energy is in the range 0.36--0.56 eV, or with Li$_{\rm Fe}^-$ to form Fe$_{\rm Li}^+$--Li$_{\rm Fe}^-$ whose formation energy is 0.51 eV \cite{Hoang2011CM} (or 0.58 eV if computed using the HSE06 functional). These results are thus consistent with the experimental observation of different defect complexes reported in the literature \cite{Chung2008,Axmann2009CM}. For comparison, the formation energy is 0.39 eV for the Co$_{\rm Li}^+$--Li$_{\rm Co}^-$ pair in LiCoPO$_4$, 0.43 eV for the Ni$_{\rm Li}^+$--Li$_{\rm Ni}^-$ pair in LiNiPO$_4$, and 0.74 eV for the Mn$_{\rm Li}^+$--Li$_{\rm Mn}^-$ pair in LiMnPO$_4$; all obtained in HSE06 calculations. The low formation energies in the case of LiCoPO$_4$ and LiNiPO$_4$, estimated to correspond to an equilibrium concentration of about 9\% for Co$_{\rm Li}^+$--Li$_{\rm Co}^-$ and 6\% for Ni$_{\rm Li}^+$--Li$_{\rm Ni}^-$ when the materials are prepared at 800$^\circ$C, are consistent with the high antisite defect concentrations often observed in experiments \cite{Allen2011JPS,Truong2013AMI,Boulineau2015CM,Devaraju2015SR,Biendicho2017IC}. It should be noted that, as demonstrated in the case of LiFePO$_4$, the antisite pair may not be the lowest-energy defect complex in the materials under {\it all} synthesis conditions. 

As reported in \cite{Hoang2011CM}, one can tune the synthesis conditions to minimize the presence of iron antisites in LiFePO$_4$ samples. The formation energy of Fe$_{\rm Li}^+$ (at $\mu_{e}^{\rm{int}}$) is highest under the most oxidizing and Li-excess environment, see figure \ref{lfp;fe}(b), specifically when the formation energies of all native defects are calculated at a point on the boundary of the stability polyhedron in the three-dimensional chemical-potential phase diagram where LiFePO$_4$ is in equilibrium with impurity phases Li$_3$Fe$_2$(PO$_4$)$_3$ and Li$_3$PO$_4$. Under these conditions, $\eta^+$ and Li$_{\rm Fe}^-$ are the dominant defects, and the formation-energy difference between Fe$_{\rm Li}^+$ and Li$_{\rm Fe}^-$ is largest \cite{Hoang2011CM}. Note that due to the exponential dependence of the concentration on the formation energy, a small formation-energy difference will result in a large difference in the defect concentrations, especially when prepared at low temperatures. The work thus predicted specific conditions under which LiFePO$_4$ with a negligible concentration of iron antisites can be obtained. The prediction was eventually confirmed by Park {\it et al.}~\cite{Park2016EES} in ``lithium-excess'' LiFePO$_4$ where Li$_{\rm Fe}^-$ was found to be the dominant defect and Fe$_{\rm Li}^+$ has a negligible concentration. The experimental work \cite{Park2016EES} also confirmed the presence of Li$_3$Fe$_2$(PO$_4$)$_3$ and Li$_3$PO$_4$ as impurity phases in certain samples when fine-tuning the synthesis conditions. The excess Li occupies the Fe sites, instead of interstitial lattice sites, which is consistent with the computational results reported in \cite{Hoang2011CM} where the dominant defect under the mentioned synthesis conditions is Li$_{\rm Fe}^0$, i.e., Li$_{\rm Fe}^-$ plus $\eta^+$. This is because lithium interstitials have a much higher formation energy than the lithium antisites \cite{Hoang2011CM}. As discussed in \cite{Park2016EES}, the lithium excess not only suppresses the iron antisites but can also open up additional Li diffusion paths perpendicular to the one-dimensional lithium channels in the material.  

\subsection{Spinel-type oxides}\label{subsec;spinel}

Truly stoichiometric LiMn$_2$O$_4$ is difficult to prepare and structural defects have been reported to occur at multiple lattice sites in this spinel oxide \cite{Tarascon1994,Xia1996,Paulsen1999,Lee2002,Martinez2014}. Martinez {\it et al.}~\cite{Martinez2014}, for example, found about 10\% of the Li ions at the Mn sites. LiMn$_2$O$_4$ samples are often made Li-overstoichiometric, either intentionally or unintentionally. The long-range charge order, expected in stoichiometric Mn$^{3+}$/Mn$^{4+}$-ordered LiMn$_2$O$_4$, is absent in Li-overstoichiometric Li$_{1+\alpha}$Mn$_{2-\alpha}$O$_4$; only short-range charge-order has been observed \cite{Sugiyama2007,Kamazawa2011}. As an electrode material, however, Li$_{1+\alpha}$Mn$_{2-\alpha}$O$_4$ shows a significantly improved cycling stability compared to stoichiometric LiMn$_2$O$_4$ \cite{Gummow1994,Xia1996}. Note that Li-overstoichiometry also occurs in the anode material LiTi$_2$O$_4$ \cite{Cava1984JSSC,Colbow1989JPS,Moshopoulou1999}.

\begin{figure}[htb!]
\centering
\includegraphics[width=9.0cm]{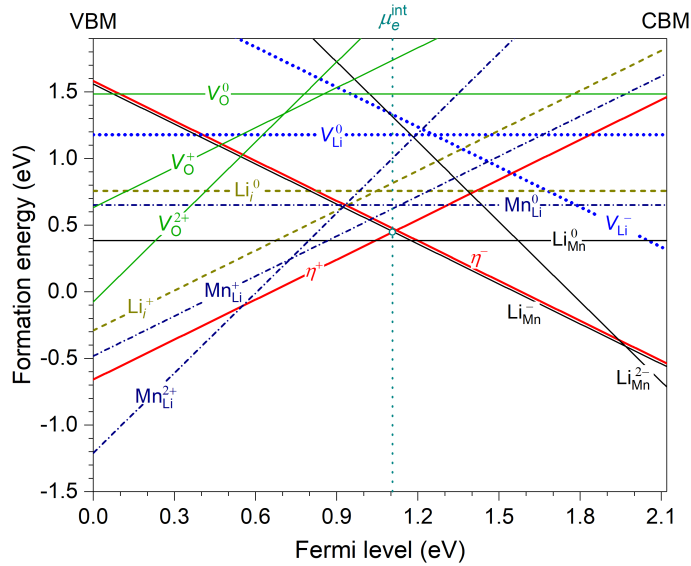}
\vspace{-0.1cm}
\caption{Formation energies of relevant native defects in LiMn$_2$O$_4$. The energies are obtained at a point in the chemical-potential phase diagram where LiMn$_2$O$_4$ is in equilibrium with Mn$_2$O$_3$ and Li$_2$MnO$_3$. Produced with data from \cite{HoangLiMn2O4}.}\label{spinel;fe}
\end{figure} 

Computationally, LiMn$_2$O$_4$ is characterized by having native defects with very low formation energies. Figure \ref{spinel;fe} shows a representative defect landscape reported in \cite{HoangLiMn2O4}. As in the other electrode materials, defects in LiMn$_2$O$_4$ include elementary defects and defect complexes, and the structure and energetics of the latter can be described in terms of those of the former. The most relevant elementary defects are $\eta^+$ (i.e., Mn$^{4+}$ at a Mn$^{3+}$ site), $\eta^-$ (high-spin Mn$^{3+}$ at a Mn$^{4+}$ site), $V_{\rm Li}^-$, Li$_i^+$, Li$_{\rm Mn}^{2-}$ (Li$^+$ at a Mn$^{3+}$ site), and Mn$_{\rm Li}^+$ (high-spin Mn$^{2+}$ at a Li site). Other charge states of the ionic defects are complexes; for example, $V_{\rm Li}^0$ is a complex of $V_{\rm Li}^-$ and $\eta^+$, Li$_{\rm Mn}^0$ (Li$_{\rm Mn}^-$) is a complex of Li$_{\rm Mn}^{2-}$ and two (one) $\eta^+$, and Mn$_{\rm Li}^0$ is a complex of Mn$_{\rm Li}^{+}$ and $\eta^-$. The Fermi level of LiMn$_2$O$_4$, $\mu_e^{\rm int}$, is predominantly determined by the dominant positively and negatively charged defects which are always $\eta^+$ and Li$_{\rm Mn}^-$. Note that LiMn$_2$O$_4$ with the Mn$^{3+}$/Mn$^{4+}$ ordering shown in figure \ref{struct}(d) was used as the host material in the calculations, where exactly half of the Mn ions are Mn$^{3+}$ and the other half are Mn$^{4+}$. The hole (electron) polaron here, $\eta^+$ ($\eta^-$), is simply an extra Mn$^{4+}$ (Mn$^{3+}$) in the supercell \cite{HoangLiMn2O4}.   

It was reported in \cite{HoangLiMn2O4} that the small polarons $\eta^+$ and $\eta^-$ in LiMn$_2$O$_4$ can occur simultaneously in the form of an electron--hole polaron pair whose formation energy is only 0.37 eV. This corresponds to an estimated equilibrium concentration of about 7\% at 800$^\circ$C (here, $N_{\rm config}=4$ for the $\eta^+$--$\eta^-$ pair); i.e., a total of about 14\% Mn$^{3+}$ and Mn$^{4+}$ ions are at the ``wrong'' Mn lattice sites with respect to the lowest-energy, Mn$^{3+}$/Mn$^{4+}$-ordered model presented in figure \ref{struct}(d). With such a high concentration, there would be strong Mn$^{3+}$/Mn$^{4+}$ disorder in real LiMn$_2$O$_4$ samples \cite{HoangLiMn2O4}. Regarding the ionic defects, the dominant defect is always the neutral complex Li$_{\rm Mn}^0$ whose formation energy is only 0.11--0.38 eV. Given this result, the synthesis of LiMn$_2$O$_4$ under equilibrium or near-equilibrium conditions is expected to result in a Li-overstoichiometric compound with a high concentration of Li at the Mn site. Its chemical formula can be written as Li[Mn$_{2-\alpha}$Li$_{\alpha}$]O$_{4}$ or, more explicitly, Li[Mn$_{1-3\alpha}^{3+}$Mn$_{1+2\alpha}^{4+}$Li$_{\alpha}^{+}$]O$_4$; here, each negatively charged lithium antisite Li$_{\rm{Mn}}^{2-}$ is charge-compensated by two hole polarons $\eta^{+}$. The average Mn oxidation state in this compound is higher than $+3.5$ because $\mu_e^{\rm int}$ is slightly on the left of the Fermi-level position where $\eta^+$ and $\eta^-$ have equal formation energies; i.e., Mn$^{4+}$ is slightly more favorable than Mn$^{3+}$. Given the Mn$^{3+}$/Mn$^{4+}$ disorder and the likely random distribution of Li$_{\rm Mn}^{0}$, long-range charge order as shown in figure \ref{struct}(d) would not be realized in real LiMn$_2$O$_4$ samples \cite{HoangLiMn2O4}, even at low temperatures, consistent with the experimental observations \cite{Sugiyama2007,Kamazawa2011}.

It was also found that the Li$^{+}$ ion at the Mn$^{3+}$ site, i.e., the lithium antisite, has a lower mobility and is unlikely to be deintercalated during charging \cite{HoangLiMn2O4}. Furthermore, Li[Mn$_{2-\alpha}$Li$_{\alpha}$]O$_{4}$ has only ($1-3\alpha$) Mn$^{3+}$ ions for the oxidation reactions (Mn$^{4+}$ is inactive because the oxidation to Mn$^{5+}$ would cost too much energy). As a result, there will be residual lithium in the fully delithiated compound, at both the Li and Mn lattice sites, i.e., Li$_{3\alpha}^{+}$[Mn$_{2-\alpha}^{4+}$Li$_{\alpha}^{+}$]O$_{4}^{2-}$. The theoretical capacity will therefore decrease from 148 mAh/g to $148(1-3\alpha)$ mAh/g. However, in spite of that capacity loss, the presence of the residual lithium in the delithiated compound can help improve the cycling stability, unlike in stoichiometric LiMn$_2$O$_4$ where the complete extraction of lithium results in the unstable $\lambda$-MnO$_{2}$ end compound \cite{Gummow1994}. The results are thus consistent with the reports that Li$_{1+\alpha}$Mn$_{2-\alpha}$O$_4$ shows an enhanced electrochemical performance \cite{Gummow1994,Xia1996}. 

\section{Delithiation mechanism and extraction voltage}\label{sec;delithiation}

In lithium-ion batteries, electrode materials are subjected to lithium extraction and (re-)insertion during charge and discharge, respectively. These processes can be regarded as corresponding to the creation of {\it electrochemically activated} defects in the electrode materials \cite{Hoang2015PRA}. For example, the delithiation reaction in a LiCoO$_2$ cathode occurs as
\begin{equation}\label{eq:oxidation}
{\rm LiCoO}_2 \rightarrow {\rm Li}_{1-x}{\rm CoO}_2 + x{\rm Li}^+ + xe^-, 
\end{equation}
where the extracted Li$^{+}$ ions then dissolve into the electrolyte and the electrons move in the opposite direction to the outer circuit. The extraction of lithium (i.e., Li$^+$ plus $e^-$) from the electrode thus corresponds to the formation of lithium vacancies ($V_{\rm Li}^0$) in LiCoO$_2$; here, the vacancies are electrochemically activated, as opposed to being activated thermally. In \cite{Hoang2015PRA}, it was demonstrated that the mechanism for delithiation and the extraction voltage can be obtained from the structure and energetics of $V_{\rm Li}^0$, respectively.

\begin{figure}[htb!]
\centering
\includegraphics[width=11cm]{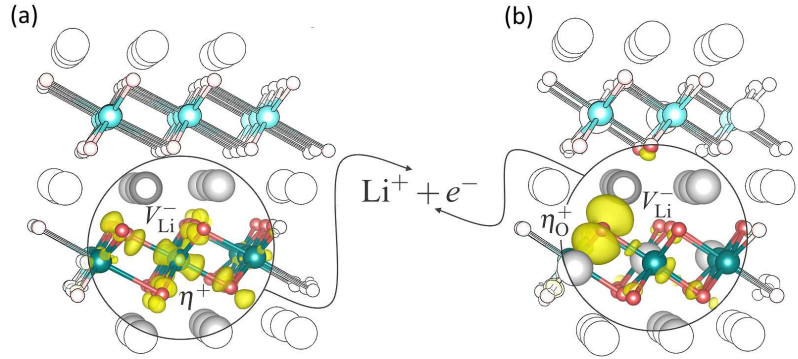}
\vspace{-0.3cm}
\caption{Mechanisms for lithium extraction in battery electrode materials, involving oxidation (a) at the transition-metal site, e.g., Mn$^{3+}$ $\rightarrow$ Mn$^{4+}$ in LiMnO$_2$ and (b) at the oxygen site, e.g., O$^{2-}$ $\rightarrow$ O$^-$ in Li$_2$MnO$_3$ \cite{Hoang2015PRA}. Large (gray) spheres are Li, medium (blue) are Mn, and small (red) are O. Charge densities associated with the hole polarons $\eta^+$ (i.e., Mn$^{4+}$) and $\eta_{\rm O}^+$ (i.e., O$^-$) are visualized as (yellow) isosurfaces.}\label{delithiation}
\end{figure} 

Figure \ref{delithiation} shows the relationship between the structure of the $V_{\rm Li}^0$ defect and the delithiation mechanism in LiMnO$_2$ and Li$_2$MnO$_3$ \cite{Hoang2015PRA}. In the vast majority of electrode materials, $V_{\rm Li}^0$ is a complex of $V_{\rm Li}^-$ and $\eta^+$, where the former is a void formed by the removal of a Li$^+$ ion from the electrode material and the latter is a hole polaron stabilized on the transition-metal ion induced by the removal of an electron from the material; see figure \ref{delithiation}(a). In this mechanism, oxidation thus occurs on the transition-metal ion and the removal of Li$^+$ ions is charge compensated by the formation of hole polarons $\eta^+$. In Li$_2$MnO$_3$, however, $V_{\rm Li}^0$ was found to be a complex of $V_{\rm Li}^-$ and $\eta_{\rm O}^+$ where the latter is an oxygen-hole polaron bound to the former; see figure \ref{delithiation}(b). In this case, the intrinsic delithiation mechanism involves the oxidation of O$^{2-}$ to O$^-$; here, charge compensation for the removal of Li$^+$ ions is provided by the formation of $\eta_{\rm O}^+$ in the material \cite{Hoang2015PRA}. 

The formation of $\eta_{\rm O}^+$ bound to $V_{\rm Li}^-$, as opposed to $\eta^+$ which is stable either in the presence or absence of other defects, has important implications for the performance of Li$_2$MnO$_3$. It is known that Li$_2$MnO$_3$ is electrochemically activated at high voltages \cite{Kalyani1999} and shows very limited electrochemical capacity \cite{Sathiya2013}. As argued in \cite{Hoang2015PRA} and \cite{Hoang2017Li2MnO3}, the difficulty in activating Li$_2$MnO$_3$ can be ascribed to the high extraction voltage and a lack of percolation pathways for electronic conduction in the bulk at the onset of delithiation when there are no or very few $V_{\rm Li}^-$. The electrochemical performance can, however, be improved via doping with electrochemically active ions \cite{Sathiya2013,Luo2016NC}. The purpose of doping is to introduce additional charge compensation and electronic conduction mechanisms highly needed at the early stages of delithiation \cite{Hoang2015PRA,Hoang2017Li2MnO3}; see also section \ref{doping;deli}.

The difference in the delithiation mechanism between, e.g., Li$_2$MnO$_3$ and LiMnO$_2$ can be traced back to the difference in their electronic structure reviewed in section \ref{sec;electronic}. This interplay between electronic structure and defect formation (and hence presence or absence and identity of active redox centers), discussed in \cite{Hoang2015PRA}, offers a simple method to computationally determine the (de)lithiation mechanism in newly discovered battery materials \cite{Yao2017CM,Yao2017CM2} where {\it a priori} knowledge of the mechanism is often lacking. The relationship between electronic structure, defect structure, and functional property will also become a basis for, e.g., high-throughput computational screening and discovery of materials with certain types of cationic and/or anionic redox centers. Note that the example presented in figure \ref{delithiation} focuses on the initial stage of delithiation. Similar calculations and analyses can always be carried out for partially delithiated compounds to understand the delithiation mechanism at the later stages, as demonstrated in \cite{Hoang2017Li2MnO3}.

Regarding the extraction voltage, the derivation presented in \cite{Hoang2015PRA} can be summarized as the following: From equation (\ref{eq;vli}), a similar expression can be written for the formation energy of arbitrarily $x$ lithium vacancies, $E^f(xV_{\mathrm{Li}}^0)$, in an electrode material. During delithiation, the lithium vacancies are electrochemically activated, i.e., 
\begin{equation}
E^f(xV_{\mathrm{Li}}^0) = 0, 
\end{equation}
assuming they readily form under the influence of an external power source with an extraction voltage $V$. Moreover, the lithium chemical potential can be expressed as 
\begin{equation}
\mu_{\rm Li}^{\ast} = E_{\rm tot}({\rm Li}) - eV, 
\end{equation}
assuming equilibrium with a metallic Li anode and the external power source which acts as a reservoir of the electrons. From these expressions, the lithium-extraction voltage can be expressed in terms of the total energies as \cite{Hoang2015PRA}
\begin{equation}\label{eq;voltage}
V = \frac{E_{\mathrm{tot}}(xV_{\mathrm{Li}}^0) - E_{\mathrm{tot}}({\mathrm{host}}) + xE_{\mathrm{tot}}({\mathrm{Li}})}{xe}.
\end{equation}
Here, $x$ can take any value to describe the lithium-content difference between any two intercalation limits, and $E_{\mathrm{tot}}({\mathrm{host}})$ can be the total energy of {\it any} starting composition chosen as the host material. In that case, $V$ is the average voltage between the two limits. Expression (\ref{eq;voltage}) is thus equivalent to that for the average voltage derived by Aydinol {\it et al}.~\cite{Aydinol1997} who considered the electrical energy caused by charge displacement. 

A similar expression for the voltage associated with lithiation can also be derived according to which the lithium-insertion process is regarded as the formation of lithium interstitials (Li$_i^0$) in the electrode material. The structure of Li$_i^0$ provides information on the lithiation mechanism \cite{HoangLiMn2O4}. Usually, one is more interested in the re-insertion of lithium into a (partially) delithiated compound. In that case, the (partially) delithiated compound should be chosen as the host material. In both delithiation and lithiation, the system is often assumed to exchange only Li$^+$ and $e^-$ with reservoirs; however, it is straightforward to include other species, if needed. The general formulation presented in \cite{Hoang2015PRA} is intuitive and can be applied to other mechanisms where lithium may not be the only extracted (inserted) species during the delithiation (lithiation) process.

\section{Electronic and ionic conduction mechanisms}\label{sec;conduction}

From the defect landscapes reported for a number of complex materials \cite{Hoang2011CM,HoangLiMn2O4,Hoang2014JMCA,Hoang2015PRA,Hoang2016CM}, including those discussed in section \ref{sec;defects}, it is clear that a change from one (nominal) defect charge state to another is associated with polaron formation, and some charged native defects have non-negative formation energies only in a small range of the Fermi-level values near midgap; see, e.g., figures \ref{lco;fe}--\ref{spinel;fe}. These features indicate that native point defects in these battery materials cannot act as sources of band-like holes and electrons, and the electronic conduction thus occurs via hopping of small polarons. The observed p-type (or n-type) thermally activated electronic conductivity reported in the literature should be understood as that associated with hole (electron) polarons. Also, the materials cannot be doped n- or p-type like a conventional semiconductor where the Fermi level can be at or very close to the band edges. Any attempt to deliberately shift the Fermi level to the VBM or CBM, e.g., via doping with impurities, will lead to spontaneous formation of native defects that counteract the effects of shifting, i.e., native defects act as charge-compensating centers \cite{Hoang2011CM}; see also section \ref{sec;doping}. Finally, the ionic conduction often occurs via migration of lithium vacancies and/or interstitials because of their higher mobility compared to other native ionic defects \cite{Hoang2011CM,HoangLiMn2O4,Hoang2014JMCA,Hoang2015PRA,Hoang2016CM}.

\subsection{Defect migration energy barrier}\label{subsec;em}

Since defect structure and energetics have been reviewed in section \ref{sec;defects}, let us now examine defect migration, particularly the migration of small polarons and lithium vacancies. We note that the methodology discussed in section \ref{sec;approach} allows defect thermodynamics and kinetics to be cleanly separated so that one can easily isolate and study a defect in detail, even when it is positively or negatively charged. For the ionic defects, the energy barrier ($E_m$) for migration can be computed using the nudged elastic band method \cite{ci-neb}. The migration of a small polaron between two positions $q_{\rm A}$ and $q_{\rm B}$, on the other hand, can be described by the transfer of its lattice distortion and the migration barrier is obtained by computing the total energies of a set of supercell configurations linearly interpolated between $q_{\rm A}$ and $q_{\rm B}$ and identifying the energy maximum \cite{Rosso2003,Iordanova122,Maxisch:2006p103}. %Iordanova123

\begin{table}[htb!]
\caption{\label{tab;em}Migration barrier $E_m$, in eV, of small hole ($\eta^+$) and electron ($\eta^-$) polarons and lithium vacancies via monovacancy ($V_{\rm Li}^-$) and divacancy ($DV_{\rm Li}^{2-}$) mechanisms.$^a$ The transition-metal ions associated with $\eta^+$ and $\eta^-$ in each compound are listed in parentheses.}
\footnotesize
\begin{tabular}{@{}llrlrlll}
\br
&\multicolumn{2}{c}{$\eta^+$}&\multicolumn{2}{c}{$\eta^-$}&$V_{\rm Li}^-$&$DV_{\rm Li}^{2-}$&Ref.\\
\mr
LiCoO$_2$&0.10&(Co$^{4+}$)&0.32&(Co$^{2+}$)&0.70&0.18&\cite{Hoang2014JMCA}\\
LiNiO$_2$&0.21, 0.28&(Ni$^{4+}$)&0.26, 0.28&(Ni$^{2+}$)&0.56, 0.66&0.26&\cite{Hoang2014JMCA}\\
LiMnO$_2$&0.39, 0.48&(Mn$^{4+}$)&0.30, 0.34&(Mn$^{2+}$)&0.58, 0.63&0.30&\cite{Hoang2015PRA}\\
NCM$_{1/3}$&0.31$^b$&(Ni$^{3+}$)&0.40$^b$&(Mn$^{3+}$)&0.64$-$0.75&0.26$-$0.28&\cite{Hoang2016CM}\\
NCA$_{1/3}$&0.23$^b$&(Ni$^{4+}$)&0.30$^b$&(Ni$^{2+}$)&0.64$-$0.95&0.35$-$0.40&\cite{Hoang2016CM}\\
Li$_2$MnO$_3$&&&0.33&(Mn$^{3+}$)&0.64, 0.82&0.29, 0.34&\cite{Hoang2015PRA}\\
LiFePO$_4$&0.22 (0.25)&(Fe$^{3+}$)&&&(0.32)&&\cite{Hoang2011CM,Johannes2012PRB}\\
LiMn$_2$O$_4$&0.46&(Mn$^{4+}$)&0.46&(Mn$^{3+}$)&0.19,$^c$ 0.47&&\cite{HoangLiMn2O4}\\
\br
\end{tabular}\\
$^a$ Results obtained in HSE06 calculations; only the values in parentheses are obtained in DFT$+U$.\\
$^b$ This work; estimated in calculations using the supercell models for LiNi$_{1/3}$Co$_{1/3}$Mn$_{1/3}$O$_2$ (NCM$_{1/3}$) and LiNi$_{1/3}$Co$_{1/3}$Al$_{1/3}$O$_2$ (NCA$_{1/3}$) and the computational details reported in \cite{Hoang2016CM}.\\ 
$^c$ Lithium can also migrate via an interstitialcy mechanism with an energy barrier of 0.12 eV or 0.49 eV. For both the vacancy and interstitialcy mechanisms, however, the lower barrier migration path is unlikely to be realized in (partially) Mn$^{3+}$/Mn$^{4+}$-disordered LiMn$_2$O$_4$; see more details in \cite{HoangLiMn2O4}.
\end{table}
\normalsize

Table~\ref{tab;em} lists the energy barriers for polaron and lithium-ion migration in select electrode materials. In these materials, lithium migration can occur via a (mono)vacancy mechanism, particularly $V_{\rm Li}^-$. The movement of the negatively charged vacancy $V_{\rm Li}^-$ in one direction is equivalent to that of a Li$^+$ ion in the opposite direction. In the layered oxides, lithium migration can also occur via a divacancy mechanism \cite{VanderVen2001,Hoang2014JMCA}, which involves a lithium divacancy $DV_{\rm Li}^{2-}$ (a pair of $V_{\rm Li}^-$). In this mechanism, the movement of one $V_{\rm Li}^-$ occurs in the presence of the other $V_{\rm Li}^-$. The monovacancy mechanism is expected to be dominant in nearly fully lithiated compounds, i.e., when the concentration of lithium vacancies is low; the divacancy mechanism is dominant in partially delithiated layered oxides where the concentration of lithium vacancies is high, i.e., when $DV_{\rm Li}^{2-}$ has a low formation energy \cite{Hoang2014JMCA,Hoang2015PRA,Hoang2016CM}. Note that the calculated migration barriers reported in the literature by other research groups may often be those for neutral vacancies $V_{\rm Li}^0$ which, as discussed in section \ref{sec;defects}, actually have two components: $V_{\rm Li}^-$ and a hole polaron.

\subsection{Conductivity and activation energy}\label{subsec;ea}

The activation energy for conduction associated with a defect can be estimated from its formation energy and migration barrier. Care should be taken, however, when making direct comparisons with the activation energy obtained in experiments. The following discussion, based on that presented in \cite{Hoang2014JMCA} and \cite{Hoang2017PRM}, is general and applicable to any materials with defect-mediated, thermally activated electronic and/or ionic conduction. 

The conductivity associated with a current-carrying defect can be defined as 
\begin{equation}\label{eq:conduct}
\sigma = q m c,
\end{equation}
where $q$, $m$, and $c$ are the defect's charge, mobility, and concentration, respectively. The mobility is assumed to be thermally activated, i.e.,
\begin{equation}\label{eq:mobility}
m=m_{0} \exp{\left(-\frac{E_{m}}{k_{\rm B}T}\right)},
\end{equation}
where $m_{0} \propto 1/T$ \cite{tilley2008defects} is a pre-exponential factor. The concentration $c$, on the other hand, can include both thermally activated and athermal defects \cite{Hoang2014JMCA,Hoang2017PRM},
\begin{equation}\label{eq:concen2}
c=c_a+c_t=c_a + c_0 \exp{\left(-\frac{E^f}{k_{\rm B}T}\right)},
\end{equation}
where $c_a$ is the athermal concentration consisting of defects that, e.g., preexist in the material before the conductivity measurements, $c_t$ is the concentration of defects that are thermally activated during the measurements at finite temperatures, and $c_{0}$ is a pre-exponential factor. In general, the $E^f$ value that enters equation (\ref{eq:concen2}) will not necessary be the same as that calculated using a specific set of conditions under which the material is prepared. This is because the experimental conditions (and hence the atomic chemical potential values) during the synthesis are generally different from those during the conductivity measurements. Exceptions are cases in which the current-carrying defects are formed via, e.g., a Frenkel or full-Schottky defect mechanism or an electron--hole polaron pair mechanism (e.g., polarons in LiNiO$_2$ and LiMn$_2$O$_4$, see section \ref{sec;defects}) and thus the defect formation energy is independent of the chemical potentials. The range of defect formation energy values calculated within the allowed range of the chemical potentials (sections \ref{sec;approach} and \ref{sec;defects}) is, however, still useful, assuming that the host compound is stable during the conductivity measurements and thus the atomic chemical potentials are still subject to the same thermodynamic constraints regarding phase stability.

From equations (\ref{eq:conduct})--(\ref{eq:concen2}), it is clear that when the athermal defects are dominant, i.e., $c_a \gg c_t$, the observed temperature-dependence of the conductivity will show an activation energy that includes only the migration barrier, i.e., 
\begin{equation}
E_{a} = E_{m}; 
\end{equation}
when the thermally activated defects are dominant, i.e., $c_t \gg c_a$, the activation energy will include both the formation energy and migration barrier, i.e., 
\begin{equation}
E_{a} = E^{f} + E_{m}.
\end{equation} 
The $c_a \gg c_t$ and $c_t \gg c_a$ situations correspond to the ``extrinsic'' (low-temperature) and ``intrinsic'' (high-temperature) regions, respectively, joined by a {\it convex} knee as often shown in Arrhenius plots of $ln(\sigma T)$ vs.~$1/T$. As far as the defect concentration is concerned, {\it the effective activation energy for conduction is thus dependent on the $c_a/c_t$ ratio}; the conductivity is, on the other hand, dependent on the total concentration ($c_a + c_t$) as seen in equation (\ref{eq:conduct}). 

Note that defect association, not included in the above equations, may have an impact on the conductivity and the activation energy at lower temperatures. In that case, the activation energy includes the migration barrier plus the energy needed to separate the defects, and in the $ln(\sigma T)$ vs.~$1/T$ plot there is usually a {\it concave} knee between the region with the associated defects and that with dissociated ones \cite{tilley2008defects}. 

In battery electrode materials, athermal defects ($c_a$) may include native defects that occur during the synthesis and get trapped in materials, including those that act as charge-compensation centers in materials doped with electrically active impurities (see section \ref{sec;doping}), and electrochemically activated positively charged (i.e., hole) polarons and negatively charged lithium vacancies in partially delithiated compounds (see section \ref{sec;delithiation}). These defects, particularly those that contribute to charge transport, can act as preexisting current-carrying defects in subsequent conductivity measurements. 

\subsection{Interpretation of conductivity data}\label{subsec;interpretation}

Let us now illustrate the earlier discussion with a few examples, using the conductivity data\footnote{In many experimental reports, the activation energy may be incorrectly derived from the slope of a $ln(\sigma)$ vs.~$1/T$ plot, instead of a $ln(\sigma T)$ vs.~$1/T$ plot as it should be for {\it thermally activated} electronic and ionic conduction [Note the pre-factor $m_{0} \propto 1/T$ in equation (\ref{eq:mobility}) for the mobility]. In that case, the {\it actual} activation energy may be larger than the reported value, usually by about a few percent.} available in the literature. The activation energy for electronic conduction was reported to be 0.11--0.16 eV in ``Li$_{1.0}$CoO$_2$'' and 0.11 eV in LiCo$_{0.97}$Mg$_{0.03}$O$_2$ \cite{Menetrier1999JMC,Levasseur2002}. Nobili {\it et al.}~\cite{Nobili2002JPCB,Nobili2005EA} found the activation energy in Li$_x$CoO$_2$ drops from about 0.4 eV for $x \sim 1$ to 0.1 eV for $x \sim 0.9$. These values, particularly those for $x<1$ (but still larger than the lithium content at which occurs the insulator--metal transition) and in the Mg-doped samples, are almost identical to the calculated migration barrier, 0.10 eV, of $\eta^+$ in LiCoO$_2$ \cite{Hoang2014JMCA}. This indicates that $c_a(\eta^+) \gg c_t(\eta^+)$ during the measurements of these samples and thus $E_{a} \sim E_{m}$. The athermal $\eta^+$ can exist as the charge-compensating defect of $V_{\rm Li}^-$ in the partially delithiated samples or of Mg$_{\rm Co}^-$ in the Mg-doped samples (see also section \ref{sec;doping}). Even the so-called ``Li$_{1.0}$CoO$_2$'' can have $c_a \gg c_t$, as indicated by the reported low activation energy \cite{Menetrier1999JMC}. Lin {\it et al.}~\cite{Lin2012}, on the other hand, found much larger activation energies, $0.97$--$1.23$ eV, in single crystals of stoichiometric LiCoO$_2$ grown by a vapor transport method. Here, the crystals must have very few preexisting $\eta^+$ defects and $\eta^+$ needs to be thermally activated during the measurements, according to which $c_t \gg c_a$. Indeed, the measured activation energies agree well with the range of values, $0.99$--$1.69$ eV, calculated from the formation energies and migration barrier of $\eta^+$ reported in \cite{Hoang2014JMCA}. The $E_{a} = 0.4$ eV value reported for some LiCoO$_2$ samples \cite{Nobili2002JPCB,Wang2018JPS} is closer to $E_m$ than to $E^f + E_m$ \cite{Hoang2014JMCA}, suggesting a high $c_a/c_t$ ratio, though it is expected to be much lower than that in some samples mentioned above. As for the ionic conduction, Wang {\it et al.}~\cite{Wang2018JPS} reported an activation energy of 0.57 eV in LiCoO$_2$, which is close to the energy barrier for lithium migration via the monovacancy mechanism \cite{Hoang2014JMCA}; see also Table \ref{tab;em}. This suggests that $c_a \gg c_t$ for $V_{\rm Li}^-$ in this case, though the total concentration ($c_a + c_t$) of $V_{\rm Li}^-$ is too small to make the divacancy mechanism favorable. Indeed, Wang {\it et al.}'s ``almost stoichiometric'' LiCoO$_2$ sample shows very low electronic and ionic conductivities \cite{Wang2018JPS}, which indicates very low concentrations of $\eta^+$ and $V_{\rm Li}^-$, especially given the fact that $\eta^+$ is highly mobile (i.e., has a very low migration barrier) \cite{Hoang2014JMCA}.

In the case of LiNiO$_2$, it should be noted there is already a significant concentration of preexisting (athermal) $\eta^+$ and $\eta^-$ defects; see section \ref{sec;layered}. In principle, both the hole and electron polarons can contribute to the electronic conductivity. However, since LiNiO$_2$ samples may often be Li-deficient, e.g, due to Li loss during the synthesis and/or when being partially delithiated, $\eta^+$ would have a higher concentration. In addition, the migration barrier of $\eta^+$ is slightly smaller than that of $\eta^-$; see Table \ref{tab;em}. As a result, $\eta^+$ is likely the majority charge-carrying species in the electronic conduction. Indeed, Seebeck coefficients were reported to be positive in LiNiO$_2$ samples (for $T<300$ K) \cite{Molenda2002}. Molenda {\it et al.}~\cite{Molenda2002} found activation energy values of $0.14$--$0.19$ eV ($\pm 0.02$ eV), which are in good agreement with the calculated migration barrier (as low as 0.21 eV) of $\eta^+$ in LiNiO$_2$ \cite{Hoang2014JMCA}. In the partially delithiated Li$_x$NiO$_2$ ($0.6 \le x < 1.0$), the activation energy only decreases slightly (which may be ascribed to the lattice parameter changes) \cite{Molenda2002}, again indicating $c_a \gg c_t$ for polarons even in the (nominal) Li$_{1.0}$NiO$_2$ sample.

The above discussion also serves as a basis for understanding LiNiO$_2$-related mixed transition-metal materials, especially Ni-rich layered oxides, where the electronic conduction is largely determined by the Ni ions. The valence-band top of these materials is predominantly composed of the Ni $3d$ states; as a result, Ni$^{2+}$ and Ni$^{3+}$ are oxidized before any other transition-metal ions (e.g., Co$^{3+}$) and small polarons associated with the Ni ions dominate the electronic transport in lithiated and partially delithiated compounds; see, e.g., the electronic structure and voltage profiles of NCM$_{1/3}$ and NCA$_{1/3}$ reported in \cite{Hoang2016CM}. Experimentally, Saadoune and Delmas \cite{Saadoune1998JSSC} reported activation energy values for electronic conduction in Li$_x$Ni$_{0.8}$Co$_{0.2}$O$_2$ varying from 0.22 eV ($x=1.0$) to 0.17 eV ($x=0.65$), in the 220--290 K range. In Li$_{1-x}$Ni$_{0.8}$Co$_{0.15}$Al$_{0.05}$O$_2$, often referred to as ``NCA'' in the literature, Amin {\it et al.}~\cite{Amin2015JES} reported a range of activation energy values, $0.22$--$0.14$ eV ($\pm 0.04$ eV) for $x =0$--$0.60$. In other materials such as Ni-rich NMC, LiNi$_{1-x-y}$Mn$_x$Co$_y$O$_2$, the activation energy was reported to be 0.24 eV (NMC811), 0.27 eV (NMC622), or 0.29 eV (NMC532) \cite{Wang2018JPS}. All these values are in good agreement with the calculated migration barrier ($0.21$--$0.31$ eV) of small polarons associated with the Ni ions in LiNiO$_2$, NCM$_{1/3}$, and NCA$_{1/3}$; see Table \ref{tab;em}. This indicates that $c_a \gg c_t$ in these samples and $E_{a} \sim E_{m}$. Amin and Chiang \cite{Amin2016JES} reported a slightly larger activation energy, 0.42 eV, for lithiated NMC532, which suggests a lower $c_a/c_t$ ratio for polarons in their sample. The activation energy, however, decreases quickly for $x>0$ \cite{Amin2016JES}, as expected. Note that, in samples with high delithiation states, $\eta^+$ associated with Co$^{4+}$ can contribute to the electronic transport; the migration barrier of this polaron in LiCoO$_2$ is only 0.10 eV \cite{Hoang2014JMCA}.

Like in the Ni-rich NMC materials, the electronic conduction in NMC333 (i.e., NCM$_{1/3}$) at high lithiation states is also characterized by that in LiNiO$_2$ due to the similarity in the nature of the valence-band top of the two compounds \cite{Hoang2014JMCA,Hoang2016CM}. Hole polarons $\eta^+$ in Li$_{1-x}$Ni$_{1/3}$Co$_{1/3}$Mn$_{1/3}$O$_2$ are predominantly Ni$^{3+}$ for $0 < x \le 1/3$, Ni$^{4+}$ for $1/3 < x \le 2/3$, or Co$^{4+}$ for $2/3 < x \le 1$, as evidenced in the voltage profile reported in \cite{Hoang2016CM}. The migration barrier of $\eta^+$ associated with Ni$^{3+}$ is calculated to be 0.31 eV. Experimentally, Wang {\it et al.}~\cite{Wang2018JPS} found an activation energy of 0.36 eV in lithiated NMC333 whereas Amin and Chiang \cite{Amin2016JES} reported a value of 0.48 eV ($\pm 0.03$ eV). The activation energy then decreases quickly as a function of $x$, from $0.29$ eV ($x=0.1$) to $0.10$ eV ($x=0.75$) \cite{Amin2016JES}, similar to what was observed in, e.g., NMC532 \cite{Amin2016JES} mentioned above. These values are comparable to the calculated migration barriers (0.21--0.31 eV) of hole polarons associated with Ni$^{3+}$ and Ni$^{4+}$ in LiNiO$_2$ \cite{Hoang2014JMCA}, NCM$_{1/3}$, and NCA$_{1/3}$ or that (0.10 eV) of hole polarons associated with Co$^{4+}$ in LiCoO$_2$ \cite{Hoang2014JMCA}; see Table \ref{tab;em}. This indicates that, in the vast majority of NMC333 samples mentioned here, $c_a(\eta^+) \gg c_t(\eta^+)$ and $E_a \sim E_m$. Also, the electronic conductivity in lithiated NMC333 is often reported to be very low \cite{Amin2016JES,Noh2013JPS} compared to that in the partially delithiated NMC333 samples, indicating a low total concentration ($c_a + c_t$) of polarons in the lithiated samples.       

Regarding the ionic conduction in the NMC and NCA materials, Wang {\it et al.}~\cite{Wang2018JPS} reported an activation energy of 0.27 eV for NMC333, in excellent agreement with the lithium migration barrier via a divacancy mechanism, $0.26$--$0.28$ eV, reported in \cite{Hoang2016CM}. The activation energies in other compounds are 0.25 eV (NMC532), 0.22 eV (NMC622), and 0.14 eV (NMC811) \cite{Wang2018JPS}, also comparable to the lithium migration barriers via a divacancy mechanism in the layered oxides as summarized in Table \ref{tab;em}. This indicates that the samples investigated by Wang {\it et al.} all have $c_a \gg c_t$ for $V_{\rm Li}^-$ and thus $E_a \sim E_m$, and $c_a + c_t$ is high enough that lithium divacancies are energetically favorable and the divacancy migration mechanism becomes dominant. The total concentration of lithium vacancies in the NMC333 sample is, however, expected to be smaller than that in the Ni-rich samples, given NMC333's much lower ionic conductivity \cite{Wang2018JPS}. Note that Amin {\it et al.}~\cite{Amin2015JES} reported a much higher activation energy, $1.25$ eV, for ionic conduction in Li$_{1-x}$Ni$_{0.8}$Co$_{0.15}$Al$_{0.05}$O$_2$ with $x=0$. In this case, the material must have $c_t \gg c_a$ for $V_{\rm Li}^-$ and the activation energy includes both the formation energy and migration barrier.

For LiFePO$_{4}$, Molenda {\it et al.}~\cite{Molenda2006SSI} reported an activation energy of 0.66 eV for the electronic conduction, comparable to that of $0.65 \pm 0.05$ eV reported by Zaghib {\it et al.}~\cite{Zaghib2007} and 0.55--0.59 eV by Amin {\it et al.}~\cite{Amin2008}. Since the calculated migration barrier of $\eta^+$ in LiFePO$_{4}$ is only 0.17--0.25 eV \cite{Ong2011PRB,Johannes2012PRB,Hoang2011CM}, the measured activation energy must include both the formation and migration energies, i.e., $\eta^+$ is thermally activated and $c_t \gg c_a$. Indeed, an estimation using the lowest calculated formation energy, 0.32 eV, of $\eta^+$ gives $E_a = 0.57$ eV \cite{Hoang2011CM}. We note that, in the ``$log(\sigma)$'' vs.~$1/T$ plot reported in \cite{Zaghib2007}, there appears to be a convex knee at the low-$T$ end which suggests the reported activation energy is in the intrinsic region, i.e., $E_a = E^f + E_m$. Regarding the ionic conduction, the activation energy was reported to be in the range 0.62--0.74 eV \cite{Molenda2006SSI,Amin2008}. These values are close to the activation energy, 0.65 eV, associated with the formation and migration of $V_{\rm Li}^-$, estimated using the lowest calculated formation energy \cite{Hoang2011CM}.   

Finally, LiMn$_2$O$_4$ is another interesting case where the concentrations of athermal polarons are already very high because (slightly more than) half of the Mn ions in the material are Mn$^{4+}$ (i.e., $\eta^+$) and the rest are Mn$^{3+}$ ($\eta^-$); see section \ref{subsec;spinel}. As a result, the activation energy for electronic conduction $E_a \sim E_m = 0.46$, the migration barrier of $\eta^+$ and $\eta^-$ \cite{HoangLiMn2O4}. The result is in excellent agreement with the experimental values, 0.40--0.44 eV, reported in the literature \cite{Massarotti1999,Iguchi2002,Fang2008}. As for the ionic conduction, the activation energies associated with the vacancy mechanism ($V_{\rm Li}^-$) in the extrinsic and intrinsic regions are 0.47 and 1.11 eV, respectively; the values associated with the interstitialcy mechanism (Li$_i^+$) are 0.49 and 1.25 eV \cite{HoangLiMn2O4}. Experimentally, Takai {\it et al.}~\cite{Takai2014} reported activation energies of 0.52 and 1.11 eV for lithium diffusion below and above 600$^{\circ}$C.

\section{Theory of doping in complex materials}\label{sec;doping}

Doping with impurities has been a widely used method to optimize the properties and performance of battery materials. First-principles defect calculations can provide a detailed understanding of its effects and identify potentially new and effective dopants. In the calculations, supercell models with different cell sizes can be used to describe a range from lightly to heavily doped materials; they may contain an isolated impurity or a complex consisting of impurities or impurities and native defects. From a materials modeling perspective, lightly doped compounds can effectively serve as model systems for understanding more complex, mixed-metal materials \cite{Hoang2012JPS,Hoang2017PRMoxides,Hoang2017Li2MnO3}. In this section, we discuss the lattice site preference and defect structure of select transition-metal and non-transition-metal impurities in a number of battery electrode materials and the effects of doping on the electronic and ionic conduction and the delithiation mechanism.  

\subsection{Lattice site preference and defect structure}\label{doping;site}

The most important task in the study of doping is to determine the lattice site preference of impurities (dopants), i.e., where the impurities are located in the host lattice when the material is prepared under certain conditions. In LiCoO$_2$, for example, a substitutional impurity X, where X is a metal, can stay at either the Li site or the Co site (The impurity is expected to be energetically unfavorable at an interstitial site). The lattice site preference can be quantified by considering the formation-energy difference \cite{Hoang2012JPS,Hoang2017PRMoxides}
\begin{equation}
\Delta E = E^f({\mathrm{X}}^{q1}_{\rm Li}) - E^f({\mathrm{X}}^{q2}_{\rm Co}),
\end{equation}
where $E^f({\mathrm{X}}^{q1}_{\rm Li})$ and $E^f({\mathrm{X}}^{q2}_{\rm Co})$ are the formation energies (at $\mu_{e}^{\rm{int}}$) of the lowest-energy configurations of X at the Li and Co sites. Here, $\Delta E > 0$ indicates that the impurity X is energetically more favorable at the Co site than the Li site, whereas $\Delta E \sim 0$ means it can be incorporated on both lattice sites with comparable concentrations. 

\begin{figure}[htb!]
\centering
\includegraphics[width=9.0cm]{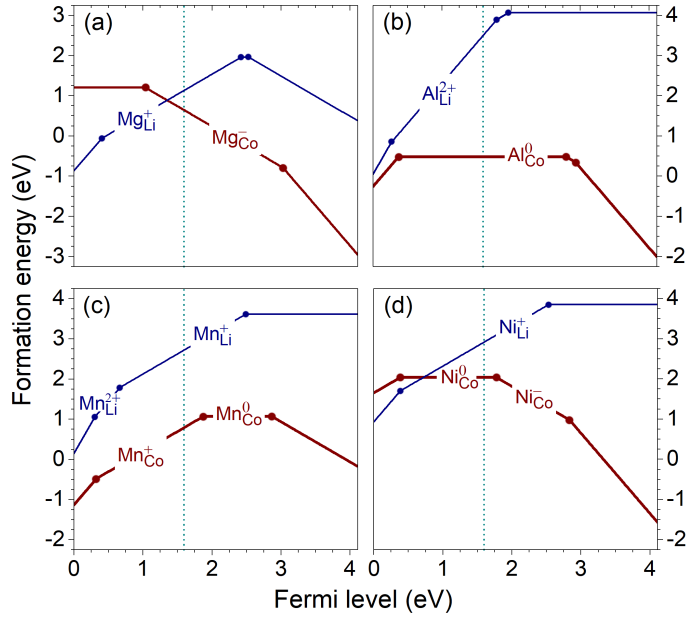}
\vspace{-0.1cm}
\caption{Relative formation energies of substitutional (a) Mg, (b) Al, (c) Mn, and (d) Ni at the Li and Co sites in LiCoO$_2$ obtained at point $M$ in the chemical-potential phase diagram (figure \ref{lco;pd}), plotted as a function of the Fermi level from the VBM to the CBM of the undoped compound. For each defect, only the true charge states are indicated. The vertical dotted line marks the Fermi level of undoped LiCoO$_2$, $\mu_{e}^{\rm{int}}$, determined by the native defects (figure \ref{lco;fe}). Produced with data from \cite{Hoang2017PRMoxides}.}\label{lco;doped}
\end{figure}

Let us illustrate with an example. Figure \ref{lco;doped} shows the formation energies of select substitutional impurities in LiCoO$_2$ for a given set of the chemical potentials \cite{Hoang2017PRMoxides}. Specifically, the results were obtained at point $M$ in the chemical-potential phase diagram where the Fermi level of the host material is at $\mu_e^{\rm int}$ (see figure \ref{lco;fe}). Under these conditions, all the impurities are energetically more favorable at the Co site, as Mg$_{\rm Co}^-$ (i.e., Mg$^{2+}$), Al$_{\rm Co}^0$ (Al$^{3+}$), Mn$_{\rm Co}^{+}$ (Mn$^{4+}$), and Ni$_{\rm Co}^0$ (low-spin Ni$^{3+}$). Some impurities are thus charged and some are neutral; the charged configurations can combine with native defects to form neutral defect complexes \cite{Hoang2017PRMoxides}. Of course, the results presented in figure \ref{lco;doped} are not the only scenario that may occur. One should note that $\mu_e^{\rm int}$ is dependent on the atomic chemical potentials of the host compound's constituent elements; as a result, the most stable charge and spin states of the impurities as well as the formation-energy difference $\Delta E$ are also dependent on the atomic chemical potentials. A systematic investigation of the impurities in LiCoO$_2$ shows that Al, Fe, and Mn are more favorable at the Co site in LiCoO$_2$, whereas Mg and Ni can be incorporated at the Co and/or Li sites depending on the synthesis conditions. For detailed results of the lattice site preference and defect structure of impurities in LiCoO$_2$, LiNiO$_2$, and LiMnO$_2$, see \cite{Hoang2017PRMoxides}. 

In another example \cite{Hoang2017Li2MnO3}, substitutional impurities Al, Fe, Mo, and Ru were found to be energetically most favorable when incorporated into Li$_2$MnO$_3$ at the Mn site, whereas Mg is most favorable at the Li sites. Interestingly, Ni can be incorporated at the Li site as Ni$_{\rm Li}^+$ (i.e., Ni$^{2+}$ at the Li ($2b$) site) and/or the Mn site as Ni$_{\rm Mn}^-$ (low-spin Ni$^{3+}$ at the Mn site), and the distribution of Ni over the lattice sites can be tuned by tuning the synthesis conditions. More interestingly, when two Ni atoms are incorporated at the Li ($2b$) site and one at the Mn site, the impurities occur as a complex of two Ni$_{\rm Li}^+$ and one Ni$_{\rm Mn}^{2-}$, i.e., all the Ni ions are in the $+2$ charge state \cite{Hoang2017Li2MnO3}. The results for Ni provide an explanation for the existence of the series of Ni-doped Li$_2$MnO$_3$ cathode materials, Li[Ni$_x$Li$_{1/3-2x/3}$Mn$_{2/3-x/3}$]O$_2$ ($0 \leq x \leq 1/2$), synthesized first by Lu {\it et al.}~\cite{Lu2001,Lu2002JES} and widely studied experimentally. This example thus illustrates how charge and spin states of a transition-metal impurity can be affected by defect--defect interaction \cite{Hoang2017Li2MnO3}.

Effects of co-doping were also investigated with particular attention paid to direct impurity--impurity interaction \cite{Hoang2017PRMoxides}. In (Ni,Mn)-doped LiCoO$_2$, for example, the dopants were found to be stable as Ni$_{\rm Co}^-$ (i.e., Ni$^{2+}$) and Mn$_{\rm Co}^+$ (Mn$^{4+}$), whereas the dopants in (Ni,Al)-doped LiCoO$_2$ are stable as Ni$_{\rm Co}^0$ (i.e., low-spin Ni$^{3+}$) and Al$_{\rm Co}^0$ (Al$^{3+}$). The Ni--Mn interaction on the Co sublattice thus leads to charge transfer between the dopants. Co-doping of LiNiO$_2$ with Co and Mn results in the formation of Co$_{\rm Ni}^0$ (i.e., low-spin Co$^{3+}$), Mn$_{\rm Ni}^+$ (Mn$^{4+}$), and $\eta^-$ (Ni$^{2+}$); in (Co,Al)-doped LiNiO$_2$, the defect structures are Co$_{\rm Ni}^0$ and Al$_{\rm Ni}^0$. Similarly, co-doping of LiMnO$_2$ with Ni and Co leads to the formation of Ni$_{\rm Mn}^-$ (i.e., Ni$^{2+}$), Co$_{\rm Mn}^0$ (low-spin Co$^{3+}$), and $\eta^+$ (Mn$^{4+}$). These co-doped systems as well as the singly doped ones discussed earlier can serve as model systems for understanding the commercially available LiNi$_x$Co$_y$Mn$_z$O$_2$ (NCM or NMC) and LiNi$_x$Co$_y$Al$_z$O$_2$ (NCA), where $x+y+z=1$ \cite{Hoang2017PRMoxides}. Nickel-rich layered oxides, for example, can be understood based on the results for the undoped and doped LiNiO$_2$ systems; see also \cite{Hoang2017PRMoxides} for a discussion of Co-rich LiCo$_{1-2x}$Ni$_x$Mn$_x$O$_2$ where the observed charge states of the impurities were explained using the results for singly doped and co-doped LiCoO$_2$. For a discussion of electronic and ionic transport in NMC and NCA materials, see section \ref{subsec;interpretation}. 

Overall, comprehensive studies of doping in battery materials reported in \cite{Hoang2012JPS}, \cite{Hoang2017PRMoxides}, and \cite{Hoang2017Li2MnO3} show that the lattice site preference of impurities does not simply depend on the ionic-radius difference between the dopant and the substituted host atom, but also on the relative abundance of the host compound's constituent elements in the synthesis environment. For transition-metal impurities, the lattice site preference also depends on the dopant's charge and spin states which are coupled strongly to the local lattice environment and can be strongly affected by the presence of co-dopant(s). These studies also provide low-energy defect models for various doped systems which are useful for experimental analyses of the materials and essential to the calculations of the electronic structure and investigations of the delithiation mechanism; see also section \ref{doping;deli}.

\subsection{Manipulation of charged native defects}

As discussed in section \ref{sec;conduction}, native defects in the electrode materials discussed here cannot act as sources of band-like carriers and the materials cannot really be doped n- or p-type like a conventional semiconductor, at least in the dilute doping limit. In \cite{Hoang2012JPS}, it is argued that rather than generating band-like carriers, ``doping'' should be understood as manipulating the concentration of native defects. As far as the defect concentration is concerned, there are two (interrelated) effects. One is the charge-compensation effect that occurs during the synthesis of the doped material. As reviewed in section \ref{doping;site}, impurities when introduced into the material can be positively or negatively charged and behave as donor-like or acceptor-like dopants, respectively. These electrically active impurities will promote the formation of native defects with the opposite charge to maintain charge neutrality. Both the impurities and the native defects get ``frozen in'' and will act as athermal, preexisting defects in subsequent material use or measurements. The second effect involves the shift of the Fermi-level position as the charge neutrality condition is re-established in the presence of the electrically active impurities; see below. 

\begin{figure}[htb!]
\centering
\includegraphics[width=9.0cm]{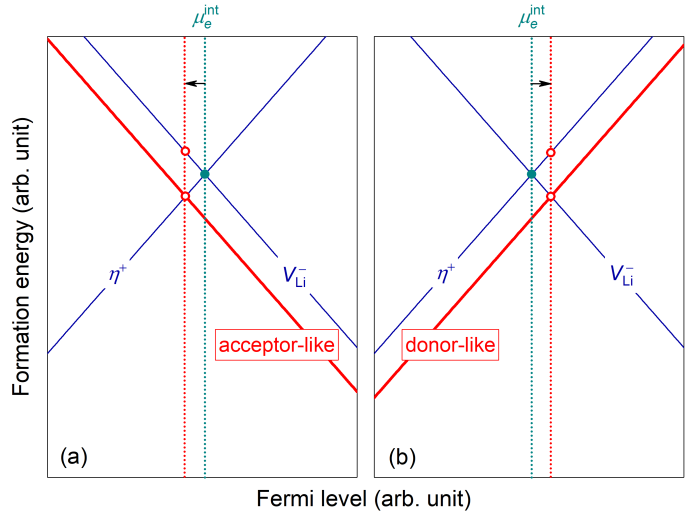}
\vspace{-0.1cm}
\caption{Schematic illustrations of (a) acceptor-like and (b) donor-like doping in a battery electrode material. $\eta^+$ and $V_{\rm Li}^-$ are assumed to be the dominant native defects which determine the Fermi level $\mu_e^{\rm int}$ of the undoped material. If incorporated with a concentration higher than that of $V_{\rm Li}^-$, an acceptor-like impurity will shift the Fermi level from $\mu_e^{\rm int}$ toward the VBM, thus decreasing (increasing) the formation energy of $\eta^+$ ($V_{\rm Li}^-$). For a donor-like impurity, the Fermi level will be shifted toward the CBM, thus decreasing (increasing) the formation energy of $V_{\rm Li}^-$ ($\eta^+$); see also \cite{Hoang2012JPS}.}
\label{doping}
\end{figure}

Figure \ref{doping} illustrates how the Fermi level gets shifted and the formation energy of $\eta^+$ and $V_{\rm Li}^-$ is modified in the case of acceptor- and donor-like doping. Because the activation energy associated with $\eta^+$ ($V_{\rm Li}^-$) in the intrinsic region contains the defect's formation energy (see section \ref{sec;conduction}), it will be decreased or increased accordingly and the change can be observed in the measured electronic (ionic) conductivity. It should be noted that, though $\eta^+$ and $V_{\rm Li}^-$ are emphasized in this example, the above analysis holds for any other positively or negatively charged native defects, including those that do not participate in mass and charge transport in the material. Similar argumentation related to Fermi-level shifting has also been developed to understand the decomposition and dehydrogenation reaction kinetics in hydrogen storage materials \cite{peles2007,hoang2009,wilson-short} and the kinetics of reduction and (re-)oxidation reactions in solid-oxide fuel cell materials \cite{Doux2018ACSAEM}.   

Let us now analyze the conductivity data for some doped battery materials in light of the above argumentation and the available computational results. Layered LiCoO$_2$ doped with Mg has been shown to enhance the electronic conductivity \cite{Levasseur2002,Nobili2005EA}. Magnesium in LiCoO$_2$ is stable as Mg$_{\rm Co}^-$ and charge-compensated by $\eta^+$ (i.e., Co$^{4+}$) \cite{Hoang2017PRMoxides}; i.e., the incorporation of the acceptor-like Mg$_{\rm Co}^-$ leads to the formation of $\eta^+$ with an equal amount. Levasseur {\it et al.}~\cite{Levasseur2002} reported an activation energy of 0.11 eV in Li$_{1.0}$Co$_{0.97}$Mg$_{0.03}$O$_2$, compared to 0.16 eV in ``Li$_{1.0}$CoO$_2$''. The value for the Mg-doped sample is almost identical to the migration barrier of $\eta^+$ (0.10 eV) \cite{Hoang2014JMCA}, indicating the concentration of preexisting $\eta^+$ defects in the doped sample is high enough so that $c_a \gg c_t$ and the activation energy contains only the migration part; see section \ref{sec;conduction}. 

The second example involves Al-doped LiFePO$_4$. Here, Al is stable as Al$_{\rm Fe}^+$, i.e., donor-like, and charge-compensated by $V_{\rm Li}^-$ \cite{Hoang2012JPS}. Amin {\it et al.}~\cite{Amin2008PCCP1} reported activation energy values of 0.15 eV and 0.65 eV (along the $c$-axis) below and above a {\it convex} knee in the ``$log(\sigma)$'' vs.~$1/T$ plot, compared to 0.55 eV in the undoped sample \cite{Amin2008}. The lower value is comparable to the migration of $\eta^+$ and this region can thus be interpreted as having $E_a \sim E_m$; i.e., $c_a \gg c_t$ at low temperatures. The higher value is higher than that for the undoped sample, which can be interpreted as $E_a = E^{f,\ast} + E_m$, where $E^{f,\ast}$ ($>E^f$) is the formation energy of $\eta^+$ that is increased due to rightward shift of the Fermi level; see figure \ref{doping}(b). Regarding the ionic conduction, Amin {\it et al.}~\cite{Amin2008PCCP2} reported values of 1.04 eV and 0.46 eV (along the $b$-axis) below and above a {\it concave} knee in the ``$log(\sigma)$'' vs.~$1/T$ plot, compared to 0.62 eV in the undoped \cite{Amin2008}. The lower value is comparable to the migration barrier (0.32 eV) of $V_{\rm Li}^-$ \cite{Hoang2011CM} and can be interpreted as being in the extrinsic region. The higher value likely includes the migration barrier plus the dissociation energy; i.e., the system is in the region where $V_{\rm Li}^-$ is bound to Al$_{\rm Fe}^+$. 

\subsection{Modification of the delithiation mechanism}\label{doping;deli}

\begin{figure}[htb!]
\centering
\includegraphics[width=9.0cm]{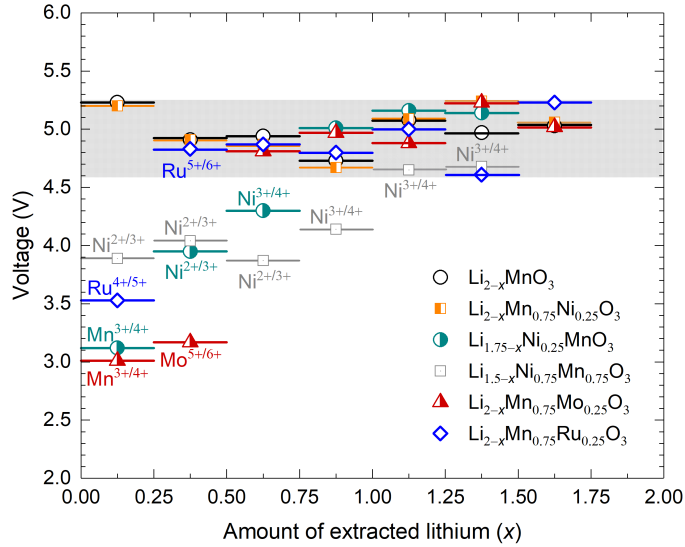}
\vspace{-0.1cm}
\caption{Voltage profiles of undoped Li$_2$MnO$_3$ and Li$_2$MnO$_3$ doped with Ni, Mo, or Ru at different lattice sites; see the text. The redox couples associated with different voltage points are indicated; unmarked voltage points in the shaded area are those associated with the redox activity on oxygen. Reprinted with permission from \cite{Hoang2017Li2MnO3}.}\label{lrm;voltage}
\end{figure}

Let us now illustrate the effects of doping on the delithiation mechanism and extraction voltage in the electrode materials by using Li$_2$MnO$_3$, a battery material with anionic electrochemical redox activity, as an example. Figure \ref{lrm;voltage} shows the calculated voltage profiles of the undoped and Li$_2$MnO$_3$ heavily doped with Ni, Mo, or Ru reported in \cite{Hoang2017Li2MnO3}. The compositions considered here include Li$_2$Mn$_{1-z}$Ni$_z$O$_3$ with Ni stable as Ni$^{4+}$ at the Mn site, i.e., Ni$_{\rm Mn}^0$; Li$_{2-z}$Ni$_z$MnO$_3$ with Ni stable as Ni$^{2+}$ at the Li ($2b$) site, i.e., Ni$_{\rm Li}^+$, and charge compensated by $\eta^-$ (i.e., Mn$^{3+}$); Li$_{2-2z}$Ni$_{3z}$Mn$_{1-z}$O$_3$ or, equivalently, Li[Ni$_y$Li$_{1/3-2y/3}$Mn$_{2/3-y/3}$]O$_2$, with one Ni at the Mn site (Ni$_{\rm Mn}^{2-}$) for every two Ni at the Li ($2b$) site, all stable as Ni$^{2+}$; Li$_2$Mn$_{1-z}$Mo$_z$O$_3$ with Mo stable as Mo$^{5+}$ at the Mn site, i.e., Mo$_{\rm Mn}^+$, and charge compensated by $\eta^-$; and Li$_2$Mn$_{1-z}$Ru$_z$O$_3$ with Ru stable as Ru$^{4+}$ at the Mn site, i.e., Ru$_{\rm Mn}^0$; $z=y/2=1/4$ in all cases. These systems were built based on the lattice site preference of the impurities determined from first principles \cite{Hoang2017Li2MnO3}. 

Clearly, doping can introduce additional delithiation mechanisms. In the examples presented in figure \ref{lrm;voltage}, they are the conventional mechanisms which involve oxidation on the transition-metal ion, as opposed to the intrinsic, oxygen-oxidation mechanism of the undoped Li$_2$MnO$_3$. The dopants are electrochemically active, except in the case of Li$_2$Mn$_{1-z}$Ni$_z$O$_3$; in some cases, the impurity--host interaction also turns some inactive Mn$^{4+}$ ions of the host into Mn$^{3+}$ which are then oxidized during lithium extraction. Charge compensation and bulk electronic conduction in the initial stages of delithiation of the doped materials are thus provided by the electrochemically active transition-metal ions. The lower voltages associated with these non-oxygen redox couples also make it easier for lithium removal. The mechanism involving the O$^{2-/-}$ redox couple is expected to be more efficient in the later stages of delithiation where the concentration of $V_{\rm Li}^-$ is high and hence bulk electronic transport via $\eta_{\rm O}^+$ may become possible \cite{Hoang2015PRA}.

Finally, it should be noted that there appears to be uncertainties and conflicting reports in the literature regarding the exact nature of the oxidized oxygen species involved in the delithiation of Li$_2$MnO$_3$-based and related materials, i.e., whether it is O$^-$ (i.e., $\eta_{\rm O}^+$ in the defect notation \cite{Hoang2015PRA}) or peroxide-like O$_2^{2-}$ \cite{Sathiya2013,Sathiya2013NM,Sathiya2015NM,Han2015JMCA,Luo2016JACS,Luo2016NC}. Tarascon and co-workers, for example, appeared to conclude that the species is O$_2^{2-}$, at least in the Ru-based systems such as Li$_2$Ru$_{1-y}$Sn$_y$O$_2$ \cite{Sathiya2013NM} and Li$_2$Ru$_{1-y}$Ti$_y$O$_2$ \cite{Sathiya2015NM}. Bruce and co-workers, on the other hand, found evidence of ``localized electron holes on oxygen'' but little or no evidence for the formation of true O$_2^{2-}$ species in Li[Li$_{0.2}$Ni$_{0.2}$Mn$_{0.6}$]O$_2$ \cite{Luo2016JACS}, i.e., Li[Ni$_y$Li$_{1/3-2y/3}$Mn$_{2/3-y/3}$]O$_2$ with $y=1/5$, and Li$_{1.2}$[Ni$_{0.13}$Co$_{0.13}$Mn$_{0.54}$]O$_2$ \cite{Luo2016NC}, which is consistent with the computational results for undoped and doped Li$_2$MnO$_3$ reported in \cite{Hoang2015PRA} and \cite{Hoang2017Li2MnO3}. The difference could be due to the strong Ru-O interaction, including the strong hybridization between the $\eta_{\rm O}^+$ $2p$ and Ru $4d$ states, as observed in Ru-doped Li$_2$MnO$_3$ compared to the Ni- or Mo-doped systems \cite{Hoang2017Li2MnO3}. It is possible that oxidized oxygen species other than or in addition to O$^-$ can form in Li$_2$MnO$_3$-related materials with a high Ru content and/or at very high degrees of delithiation where the lattice environment can be drastically different from the layered structure of the lightly delithiated compounds. Comparative experimental studies of, e.g., Ni- or Mo-doped vs.~Ru-doped Li$_2$MnO$_3$ or similar systems could shed more light on this issue.

\section{Conclusions and outlook}\label{sec;conclusions}

We have described a theoretical framework based on defect physics that is able to provide a detailed understanding of complex energy materials. Through examples involving complex oxide and polyanionic compounds, we have demonstrated that the approach is effective in predicting defect landscapes under different synthesis conditions, providing guidelines for defect characterization and defect-controlled synthesis, uncovering the mechanisms for electronic and ionic conduction and electrochemical extraction and (re-)insertion, and providing an understanding of the effects of doping. The materials were also found to exhibit a rich defect physics resulting from the ability of the transition-metal ions to exist in different charge and spin states and the strong coupling between charge, spin, and local atomic structure. Although the examples involve battery materials, the approach and the underlying physical principles discussed here are applicable to other classes of materials. In general, the approach can be applied to any materials in which the defect physics drives the properties of interest.

The power of the theoretical and computational approach described in this review comes in part from systematic first-principles defect studies. These studies, in turn, could benefit from a certain level of automation, from setting up the calculations to processing the results. The energy landscapes in materials such as those for battery electrodes are, however, very complex and the system containing defects can easily be trapped in a local minimum. The computational tools and algorithms, therefore, have to be very robust to ensure that the obtained defect configurations are indeed the ground state. The data generated from such high-throughput first-principles defect calculations could then be used in the construction of an interactive database for defects in solids and in further studies using data-driven approaches to materials screening and discovery. 

The equilibrium conditions with impurity phases often assumed in the calculations provide experimentalists with approximate synthesis conditions under which a certain defect landscape is obtained. Defect calculations, on the other hand, could also benefit from additional information from experiments regarding the actual conditions during synthesis, including the possible formation of impurity and/or intermediate phases. It is usually difficult to know the exact experimental conditions to map them onto specific points in the phase diagram at which the formation energies are calculated. Also, a full exploration of high-dimensional chemical-potential phase diagrams, as required for multinary host compounds, may not always be practical. Information about impurity phases or metastable intermediates obtained in, e.g., {\it in situ} x-ray diffraction \cite{Shoemaker2014PNAS}, therefore, could be useful in helping determine more accurately experimentally relevant areas in the phase diagram. Such joint efforts between theory and experiment would benefit further the understanding and design of complex functional materials.

\section*{Acknowledgement}
\addcontentsline{toc}{section}{\protect\numberline{}Acknowledgement}%

This work made use of computing resources at the Center for Computationally Assisted Science and Technology at North Dakota State University. MDJ acknowledges funding for this project by the Office of Naval Research (ONR) through the Naval Research Laboratory’s Basic Research Program.

\section*{References}
\addcontentsline{toc}{section}{\protect\numberline{}References}%
%\bibliography{batterymaterials}

\end{document}